\documentclass[12pt, a4paper]{article}
\usepackage[margin=2.5cm]{geometry} 

\usepackage{graphicx}
\usepackage{eurosym}
\usepackage{caption}
\usepackage{subcaption}
\usepackage{longtable} 
\usepackage[usenames]{color}
\usepackage{amsmath}
\usepackage[table]{xcolor}
\usepackage{comment}
\usepackage[flushleft]{threeparttable}
\usepackage{float}
\usepackage[title]{appendix}
\usepackage[normalem]{ulem}
\usepackage{booktabs} 

\usepackage[square,numbers]{natbib}
\usepackage[colorlinks=true, citecolor=blue, linkcolor=black, urlcolor=blue]{hyperref}

\begin{document}

\begin{center}
{\LARGE{A distributional modelling approach with application to electricity price forecasting}}

\par\end{center}

\begin{center}
{\LARGE{}\vspace{0.5cm}
 }{\large{Ciarreta, A.$^{a}$, Muniain, P.$^{b}$ and Zarraga,
A.$^{c}$} }
\par\end{center}

\vspace{0.3cm}

{\footnotesize{$^{a}$ Department of Economic Analysis, Euskal Herriko Unibertsitatea (EHU). Avda. Lehendakari Aguirre, 83. 48015
Bilbao. Spain. E-mail: aitor.ciarreta@ehu.eus.} }{\footnotesize\par}

{\footnotesize{$^{b}$ Department of Applied Mathematics, Euskal Herriko Unibertsitatea (EHU). Torres Quevedo Ingeniaria Plaza, 1. 48013 Bilbao.
Spain. E-mail: peru.muniain@ehu.eus.} }{\footnotesize\par}

{\footnotesize{$^{c}$ (Corresponding author) Department of Quantitative Methods, Euskal Herriko Unibertsitatea (EHU). Avda. Lehendakari Aguirre, 83. 48015
Bilbao. Spain. E-mail: ainhoa.zarraga@ehu.eus.} }{\footnotesize\par}

\vspace{0.5cm}

\begin{abstract}  
The increasing volatility of electricity prices driven by renewable energy integration, market shocks, and regulatory changes has reinforced the need for forecasting methods that go beyond point predictions and accurately describe the full conditional price distribution. This paper applies the Generalised Additive Models for Location, Scale and Shape (GAMLSS) framework to forecast Spanish day-ahead electricity prices using hourly data from 2020 to 2024. Alternative specifications based on Normal, Johnson's SU (JSU), and Sinh–Arcsinh (SHASH) distributions are considered, allowing the location, scale,  and shape parameters to vary with market fundamentals, including electricity demand, renewable generation, seasonal effects, and regulatory and geopolitical risk factors. Forecasts are generated using a rolling-window approach and evaluated through the mean absolute error (MAE), pinball loss, and Diebold–Mariano tests. The results show that flexible distributional specifications improve forecasting performance relative to a naive benchmark and the standard normal specification. While SHASH and JSU specifications provide the lowest point forecasting errors, the hourly analysis reveals substantial intraday variation in relative performance across specifications. JSU specification with all four parameters driven by covariates achieves the best probabilistic forecasting performance, particularly in the tails of the distribution. Diebold–Mariano tests confirm the statistical significance of these improvements. These findings highlight the importance of modelling time-varying shape distributional parameters and demonstrate the value of GAMLSS models for forecasting and risk management in increasingly volatile electricity markets.
\end{abstract}

\vspace{0.5cm}


\vspace{0.5cm}


\pagebreak

\section{Introduction}

In recent years, electricity price forecasting has become increasingly important due to the sharp rise in volatility in European wholesale markets. The 2022 energy crisis, together with the growing penetration of renewable energy sources, intensified the debate on the need to reform electricity market design in order to mitigate the risks that high volatility represents for market participants. Predictable electricity prices are essential for households, firms, and policy makers, as they require stability for short-term operational and consumption decisions, as well as clear signals for long-term investment planning.

The increase in market uncertainty has reinforced the need for accurate forecasting tools to support risk management and hedging strategies. Financial instruments based on futures contracts are among the mechanisms that the European Commission seeks to encourage through its proposed electricity market reform \citep{EC_EMD_2023}.
However, the effectiveness of such instruments critically depends on the availability of reliable electricity price forecasts. Long-term contracts, regulated tariffs, and hedging products require accurate assessments of future market conditions. Consequently, electricity price forecasting has become a central topic in energy economics literature and an important component of the ongoing discussion on electricity market reform and risk mitigation.

Forecasting electricity prices is particularly challenging due to their several distinctive characteristics from conventional financial assets. Electricity is a non-storable commodity at large scale, and its prices typically display strong seasonality, mean reversion, abrupt spikes, volatility clustering, negative prices, asymmetry, and heavy tails. These characteristics have become more pronounced with the growing integration of intermittent renewable energy sources such as wind and solar power, whose variability introduces additional uncertainty into market outcomes.

Accordingly, there is extensive literature on electricity price forecasting that can be categorised according to the methodology employed (see \cite{Weron2014} and \cite{Lago2021} for comprehensive reviews). Traditional econometric methods primarily rely on univariate and multivariate time-series regression models to capture the main characteristics of electricity prices. Widely used methods include autoregressive integrated moving average (ARIMA) models, autoregressive models with exogenous variables (ARX), vector auroregressive (VAR) models, and their extensions. These approaches have been shown to be effective for short-term forecasting\footnote{For early contributions, see \cite{Contreras2003} and \cite{Aggarwal2009}.}. However, standard linear time-series specifications usually assume Gaussian errors and constant conditional variance. These assumptions are frequently not met in electricity markets. 

More advanced modelling frameworks have incorporated time-varying volatility structures and nonlinear dynamics to address these limitations. Gneralized autoregressive conditional heteroskedasticity (GARCH) models -- at times in combination with ARIMA models or regime-switching approaches-- are an example (see for instance \cite{Tan2010} and \cite{Cifter2013}). However, these approaches primarily focus on modelling the conditional mean and variance of electricity prices and may still face difficulties to capture the full conditional distribution of electricity prices, particularly asymmetry and heavy tails.

In parallel, other studies have employed machine learning methods to forecast electricity prices given their ability to capture complex nonlinear relationships among market variables. Examples include support vector regression \citep{Pai2005}, random forests and gradient boosting methods \citep{Lago2018}, and a variety of deep learning techniques (see \cite{Weron2014}, \cite{Nowotarski2018}, and \cite{Lago2021}). These methods have shown promising forecasting performance, particularly in short-term applications involving large datasets and highly nonlinear interactions among explanatory variables. However, many studies employing machine learning methods focus primarily on point forecasting rather than modelling the complete conditional distribution of electricity prices.

This limitation becomes particularly relevant in electricity markets, where the growing penetration of renewable energy sources has led to more frequent price spikes, negative prices, and heavy‑tailed distributions. In such environments, expected prices alone provide limited information for market participants because they are also exposed to asymmetric risks and extreme price events. Consequently, forecasting approaches capable of modelling the full conditional distribution of prices may provide a richer and more informative representation of market uncertainty.

Generalized Additive Models for Location, Scale and Shape (GAMLSS), introduced by \cite{Rigby2005}, offer a flexible framework for modelling the entire conditional distribution of electricity prices. Unlike standard econometric or machine‑learning approaches, GAMLSS allows not only the conditional mean but also other parameters of the response distribution -- including variance, skewness, and kurtosis -- to depend on explanatory variables through linear or nonlinear functions. This flexibility makes GAMLSS particularly well suited to capturing the behaviour of electricity prices. 

Several studies have applied GAMLSS to electricity price forecasting. Some examples include the early contribution of \cite{Serinaldi2011}, who models the dynamically varying distribution of electricity prices in the California and the Italian Power Exchange markets. More recent studies have explored GAMLSS and related distributional approaches for probabilistic forecasting in European electricity markets (see \cite{HirschZiel2024} for an application to the German intraday market). Other contributions have emphasised the importance of modelling the entire conditional distribution of electricity prices \citep{Nowotarski2018,NarajewskiZiel2020}. These studies suggest that flexible distributional frameworks can better capture market uncertainty and improve forecasting performance under highly volatile market conditions.


Nevertheless, the application of GAMLSS models to the Iberian electricity market remains largely unexplored. The Spanish zone of the Iberian market represents a particularly relevant case due to its increasing penetration of renewable generation, especially wind and solar power, and the substantial price volatility observed in recent years. These characteristics make the market especially suitable for assessing forecasting approaches capable of modelling time-varying distributional properties.

Therefore, this paper contributes to the literature by applying GAMLSS models to forecast day-ahead electricity prices in the Spanish electricity market. Specifically, the proposed approach models the entire conditional distribution of electricity prices by allowing its parameters to vary with exogenous variables. In particular, the main contributions of the paper are as follows:

\begin{itemize}
\item Use of flexible distributional specifications to model dynamically the mean, variance, skewness, and the kurtosis of the Spanish day-ahead electricity prices.

\item To incorporate exogenous variables related to market fundamentals into the dynamic specification of distribution parameters.

\item To evaluate both point forecasts and probabilistic forecasts.
\end{itemize}

The remainder of the paper is organised as follows. Section \ref{method} presents the methodology, including the GAMLSS framework and the forecasting performance measures used in the analysis. Section \ref{applic} describes the model specification and the data used in the empirical application. Section \ref{results} presents and discusses the empirical results. Finally, Section \ref{conclusions} concludes.

\section{Methodology} \label{method}

This section describes the methodology used. Section \ref{theory} provides an introduction to GAMLSS models. Section \ref{specification} details the distributions considered for the response variable. Section \ref{forecast} explains the methodology used to forecast prices and the selection criteria.

\subsection{GAMLSS theory}\label{theory}

Generalized Additive Models for Location, Scale and Shape (GAMLSS), introduced by \cite{Rigby2005}, are a flexible statistical modelling framework. It extends Generalized Linear Models (GLM) and Generalized Additive Models (GAM), originally formulated by \cite{Nelder1972} and \cite{Hastie1990}, respectively, by relaxing the exponential family distribution assumed for the response variable, $Y$, and allowing multiple parameters of the distribution (not only the mean) to be modelled as parametric and/or additive non-parametric (smooth) functions of explanatory variables, $X$, and/or random effects terms. The distribution is usually parameterised in terms of location ($\mu$), scale ($\sigma$), and shape parameters, such as skewness ($\nu$) and kurtosis ($\tau$). 

The response variable, $Y$, is therefore modelled using a  distribution function, characterised by up to four parameters ($\mu$, $\sigma$, $\nu$, and $\tau$). The GAMLSS model can be written as:\footnote{See \cite{Rigby2020} for a detailed explanation.}

\begin{equation*}
Y \sim \mathcal{D} (\mu, \sigma, \nu, \tau) \\
\end{equation*}

\begin{equation*}
\eta_1 = g_1(\mu) = X_1\beta_1 + s_{11}(x_{11})+\ldots +s_{1J_1}(x_{1J_1})
\end{equation*}

\begin{equation}
\eta_2 = g_2(\sigma) = X_2\beta_2 + s_{21}(x_{21})+\ldots +s_{2J_2}(x_{2J_2})
\end{equation}

\begin{equation*}
\eta_3 = g_3(\nu) = X_3\beta_3 + s_{31}(x_{31})+\ldots +s_{3J_3}(x_{3J_3})
\end{equation*}

\begin{equation*}
\eta_4 = g_4(\tau) = X_4\beta_4 + s_{41}(x_{41})+\ldots +s_{4J_4}(x_{4J_4}),
\end{equation*}

\noindent where $\mathcal{D}$ is the response variable distribution with up to four parameters; $g_i()$ for $i=1,\ldots,4$ is the link function to model the $i$th parameter of the distribution; and $s_{ij}(x_{ij})$ for $j=1,\ldots,J_i$ and $i=1,\ldots,4$ are smoothing functions for explanatory variables.


The response variable distribution $\mathcal{D}$ encompasses a  wide range of distributions, including continuous distributions exhibiting high skewness and kurtosis, and  discrete distributions. 

\subsection{GAMLSS specification}\label{specification}

We consider several probability distribution functions for the response variable that might be appropriate for the data. However, given computational constraints resulting in non-convergence for several candidates, we restrict our analysis to the following continuous distributions:\footnote{See \cite{Rigby2020} for a detailed explanation.} 
\begin{itemize}
\item Normal (NO) 

The normal distribution is taken as the benchmark. It is a two-parameter distribution (mean and variance). Its density function is:

\begin{equation} \label{eq:no}
f(y|\mu,\sigma) = \frac{1}{\sigma \sqrt{2\pi}} \exp\left( -\frac{(y - \mu)^2}{2\sigma^2} \right)
\end{equation}

\noindent for $y \in (-\infty,\infty)$, where $\mu \in (-\infty,\infty)$ is the mean (location parameter), and $\sigma >0 $ is the  standard deviation (scale parameter).

\item Johnson's SU (JSU) 

The Johnson's SU (JSU) is a four-parameter distribution proposed by \cite{Johnson1949} as an inverse hyperbolic sine function transformation of a normally distributed variable. It is very flexible and extends the normal distribution to capture high asymmetry and kurtosis. The density function takes the form:

\begin{equation}
f(y|\mu,\sigma, \nu, \tau) = 
\frac{\tau}{\sqrt{2\pi}c\sigma(s^2+1)^{1/2}} \exp \left[ -\frac{1}{2} z^2 \right]
\end{equation}


\noindent for $y \in (-\infty, \infty)$, where $\mu$ and $\sigma$ are defined as in Equation (\ref{eq:no}), $\nu \in(-\infty,\infty)$ is the skewness parameter ($\nu >0$ indicates a positively skewed distribution and $\nu <0$ indicates negative skewness), $\tau>0$ is the shape parameter that affects the kurtosis (as $\tau \to \infty$ the distribution converges to the normal) , and where

\begin{equation}
z = -\nu + \tau sinh^{-1}(s) = -\nu + \tau ln \left[ s + (s^2+1)^{1/2} \right]
\end{equation}

\begin{equation}
s=\frac{y-\mu + c\sigma \omega^{1/2} sinh(\nu/\tau)}{c\sigma}
\end{equation}

\begin{equation}
c=\left[\frac{1}{2}(\omega - 1)[\omega cosh(2\nu/\tau) + 1] \right ]^{-1/2}
\end{equation}

\begin{equation}
\omega = exp(1/\tau^2)
\end{equation}

\item Sinh–Arcsinh (SHASH) 

The Sinh--Arcsinh (SHASH) distribution is a family of four-parameter distributions introduced by \cite{Jones2009} that applies the sinh-arcsinh transformation to a generating distribution with only two parameters, usually the normal. It captures asymmetry and tails that are lighter or heavier than the normal distribution. The probability density function is given by:

\begin{equation}
f(y|\mu,\sigma, \nu, \tau) = 
\frac{c}{\sqrt{2\pi}\sigma(1+z^2)^{1/2}} \exp (-r^2/2)
\end{equation}

\noindent where 

\begin{equation}
r = \frac{1}{2} \left[ \exp [ \tau sinh^{-1}(z)] - \exp [-\nu sinh^{-1}(z)] \right ]
\end{equation}

\begin{equation}
c = \frac{1}{2} \left[ \tau \exp [ \tau sinh^{-1}(z)] + \nu \exp [-\nu sinh^{-1}(z)] \right ]
\end{equation}

\begin{equation}
z = \frac{y- \mu}{\sigma} 
\end{equation}

\noindent for $y \in (-\infty, \infty)$, where $\mu \in (-\infty, \infty)$ is the median, $\sigma$ is defined as in Equation (\ref{eq:no}), and $\nu >0$, and $\tau>0$ are shape parameters. $\nu<1$ provides left tail heavier than the normal and $\nu>1$ lighter, and $\tau$ affects the right tail in a similar way.
\end{itemize}

We specify the identity link function $g_1()$ and the logarithmic link function $g_2()$ for each of the three candidate probability distribution functions. The location and scale parameters are modelled as functions of a set of explanatory variables, $X$, detailed in Section \ref{data}.

With respect to the shape parameters, the identity and the logarithmic link functions $g_3()$ are considered for JSU and SHASH distributions, respectively. Additionally, the logarithmic link function $g_4()$ is employed for both distributions.

For the JSU distribution, three alternative specifications are considered for the shape parameters:

\begin{itemize}
\item Both shape parameters are assumed to be constant.

\item The kurtosis parameter is assumed to be constant, while the asymmetry parameter is specified as a function of the explanatory variables.

\item Both shape parameters are specified as functions of the explanatory variables.

\end{itemize}

For the SHASH distribution, both shape parameters are assumed to be constant.

Therefore, a total of five different GAMLSS model specifications are considered, summarised in Table \ref{tab:link}, taking into account the three candidate distributions and the alternative link function specifications.

\begin{table}[H]
\centering \caption{Link functions} \label{tab:link}
\begin{threeparttable}
\begin{tabular}{cccccc}
  \hline
Function & NO & JSU(2) & JSU(3) & JSU(4) & SHASH(2) \\ 
  \hline
  $g_1(\mu)=\mu$ & $X\beta_1$ & $X\beta_1$ & $X\beta_1$ &  $X\beta_1$ & $X\beta_1$ \\
  $g_2(\sigma)=log(\sigma)$ & $X\beta_2$ & $X\beta_2$ & $X\beta_2$ &  $X\beta_2$ & $X\beta_2$ \\
  $g_3(\nu)$ & --- & $\nu= \beta_{\nu}$ & $\nu=X\beta_3$ &  $\nu=X\beta_3$ & $log(\nu)=\beta_{\nu}$ \\
  $g_4(\tau)=log(\tau)$ & --- & $\beta_{\tau}$ & $\beta_{\tau}$ &  $X\beta_4$ & $\beta_{\tau}$ \\
   \hline
\end{tabular}
\begin{tablenotes}
\footnotesize
\item Note: NO, JSU, and SHASH denote the Normal, Johnson's SU, and Sinh–Arcsinh distributions, respectively. The number in parentheses indicates how many distribution moments are modelled as non-constant, i.e., as functions of the explanatory variables. For example, JSU(2) means that the first two moments are non-constant. $X$ is the set of explanatory variables described in Section \ref{data}.
\end{tablenotes}
\end{threeparttable}
\end{table}


\subsection{Price forecasting and error measures}\label{forecast}
We use a rolling window of 730 observations (two years) to estimate the model $Y = X\beta + \epsilon$, where $X$ denotes the set of explanatory variables and $Y$ the hourly electricity price (both described in Section \ref{data}). Due to the large number of regressors in the model, we employ the lasso estimation method proposed by \cite{Tibshirani1996}. This approach combines features of subset selection and ridge regression to perform variable selection by shrinking some coefficients exactly to zero and thereby selecting a parsimonious subset of regressors. We then obtain one-day-ahead forecasts of the price and iterate this procedure until forecasts for the final three years of the sample period are obtained. We conduct the analysis on each of the 24 hourly series separately.

We assess the point forecasts accuracy using the popular Mean Absolute Error (MAE), defined for each hour $h$ as:
\begin{equation}
MAE_h= \frac{1}{m}\sum_{t=1}^m |Y_{t,h} - \hat{Y}_{t,h}| \hspace{0.8cm} h=1, \ldots, 24
\end{equation}

\noindent where $\hat{Y}_{t,h}$ denotes the one-day-ahead forecast of the dependent variable and $m$ is the number of forecasts. We also compute the average MAE across all hours. Lower values of the MAE indicate superior forecasting performance.

Electricity price forecasting has progressively shifted from point predictions
towards probabilistic approaches that characterise the full conditional
distribution of future prices \citep{Weron2014, Nowotarski2018}. Accordingly, this paper also considers probabilistic forecasts, which are essential for risk-aware bidding strategies, portfolio optimisation and reserve procurement. In these contexts, decision-makers require not only a central estimate, but also quantitative uncertainty bounds \citep{Hong2016}.

To evaluate probabilistic forecasts, we adopt the pinball loss function (also known as quantile or tick loss), which has been used by \cite{MuniainZiel2020} and \cite{Lago2021}, among others, in electricity price forecasting. The pinball loss function for a single observation at hour $h$ is defined as: 

\begin{equation}
    \mathcal{L}_{t,\tau,h}(Y_{t,h}, \hat{Y}_{t,\tau,h})=
        \begin{cases}
        (1-\tau) \,\bigl(\hat{Y}_{t,\tau,h} - Y_{t,h}\bigr)
            & \text{if } \hat{Y}_{t,\tau,h} \geq Y_{t,h} \\[6pt]
        \tau \,\bigl(Y_{t,h} - \hat{Y}_{t,\tau,h} \bigr)
            & \text{if } \hat{Y}_{t,\tau,h} < Y_{t,h}
    \end{cases}
    \label{eq:pinball}
\end{equation}

\noindent where $\hat{Y}_{t, \tau, h}$ denotes the one-day-ahead forecast of the dependent variable at the $\tau$-quantile, with $\tau \in (0,1)$. The pinball loss function is asymmetric. For instance, underprediction at the 0.9 quantile is penalised nine times more heavily than overprediction. It is precisely this asymmetry that incentivises the model to reproduce the correct conditional quantile of the price distribution.

We calculate the pinball score of quantile $\tau$ and hour $h$ as:

\begin{equation}
PB_{\tau,h} = \frac{1}{m} \sum_{t=1}^m \mathcal{L}_{t,\tau,h}(Y_{t,h}, \hat{Y}_{t,\tau,h})
\end{equation}

We consider the quantiles $\tau=0.01, 0.05, 0.95$, and $0.99$ in order to assess forecast accuracy in both lower and upper tails of the conditional price distribution. 
A lower value of $PB_{\tau,h}$ indicates better forecast accuracy at quantile $\tau$. As in the case of the MAE, we additionally compute the average pinball score across all hours.

Finally, we compare the price forecasting performance of the model specifications using the Diebold and Mariano (DM) test \citep{diebold1995paring}. The DM test assesses the null hypothesis of equal predictive accuracy between two competing forecasts based on a loss differential series. Its key advantage is its generality: it does not require nested models, imposes no assumptions on the forecasting model, and accommodates a broad class of loss functions. We adopt the MAE, which is robust to the price spikes characteristic of electricity markets, for point forecasts. Pinball loss is used for probabilistic forecasts; it captures the asymmetric cost of quantile violations by penalising under- and over-prediction differently according to the target quantile. 

However, the DM test is known to exhibit size distortions in finite samples, tending to reject the null hypothesis too frequently when the evaluation window is short. To address this, we apply the small-sample correction proposed by \cite{harvey1997testing}, which rescales the DM statistic by a factor of $\sqrt{(T+1)/T}$ for one-step-ahead forecasts and uses a $t_{T-1}$ distribution for inference rather than the standard normal. This adjustment yields more reliable test sizes for samples lengths typical in electricity price forecasting applications. The forecast horizon is always one since all forecasts are produced at a daily horizon for each hour of the day independently. Consequently, the long-run variance estimator simplifies to the sample variance of the loss differential series, with no autocorrelation correction required.

\section{Application} \label{applic}

This section is divided into two parts. Section \ref{data} describes the data used in the empirical analysis and the model specification, while Section \ref{statistics} presents the main descriptive statistics.

\subsection{Data and model specification}\label{data}

We use prices ($P$) from the Spanish zone of the Iberian Day-Ahead electricity market as the response variable, covering the period from 1 January 2020 to 31 December 2024. The data comprise $43,848$ hourly observations over $1,827$ days. We select this market because it has experienced fluctuations linked to external shocks and regulatory changes that have not occurred in other neighbouring markets during the analysed years. 

Specifically, several events had an impact on the market's performance during the period of study. Firstly, Spain implemented a nationwide lockdown from 15 March to 21 June 2020. This period was characterised by a drastic fall in the economic activity, resulting in a decline and low variability of electricity prices. After the lockdown had been lifted, mobility restrictions remained in place. Secondly, a regulatory change affected the bidding price limits. On 6 May 2021, the National Commission for Markets and Competition, approved new operating rules for daily and intraday trading, effective as from 7 July 2021\footnote{See BOE-A-2021-8362 [Spanish Official Gazette] for more information. https://www.boe.es/boe/dias/2021/05/20/pdfs/BOE-A-2021-8362.pdf}. These rules adapt the supply limits to the European matching limits that came into effect. The upper limit increased from $180.3$€/MWh to $4000$€/MWh (it was initially $3000$€/MWh, but this changed on May 10, 2022), and the lower limit dropped from $0$€/MWh to $-500$€/MWh in July 2021. Thirdly, the beginning of the war in Ukraine on 24 February 2022 caused a sharp rise in prices and volatility due to uncertainty. Finally, the so-called Iberian exception, which is a market-specific regulatory measure, came into force on 15 June 2022 with the aim of limiting the price of gas on the wholesale market and addressing the economic consequences of rising energy prices.

These events and regulatory changes are taken into account when selecting the explanatory variables for modelling the different moments of the distributions of the response variable. Additionally, several dummy variables, as well as forecasts of demand and supply characteristics, are considered. A detailed description of the set of explanatory variables, $X$, is provided below:\footnote{Data are available from the European Network of Transmission System Operators for Electricity (ENTSO-E) transparency platform, https://transparency.entsoe.eu/.}

\begin{itemize}
    \item $W$: Day-ahead generation forecasts for hourly wind generation (MWh). It includes on-shore and off-shore wind generation.\footnote{We used interpolation to replace the missing observation on 5 January 2020.}
    \item $S$: Day-ahead generation forecasts for hourly solar generation (MWh).
    \item $LF$: Day-ahead total hourly load forecast (MW).
    \item $D1, D2, \ldots, D7$: Dummy variables for the days of the week.
    \item $M1, M2, \ldots, M12$: Dummy variables for the months of the year.
    \item $DG$: Dummy variable with a value of 1 for the period of the Iberian exception for the gas price (from June 15, 2022 onward) and 0 otherwise.
    \item $DP$: Dummy variable with a value of 1 from the day on which the maximum and minimum price limits were changed from 0 and 180.3\euro /MWh to $-500$ and $3000$\euro /MWh, respectively (from 7 July $2021$ onward).\footnote{See BOE-A-2021-8362 [Spanish Official Gazette] (https://www.boe.es/boe/dias/2021/05/20/pdfs/BOE-A-2021-8362.pdf). In May 10, 2022 the maximum limit was raised to 4000\euro /MWh. However, this change has not been relevant in terms of prices, so we do not consider an additional dummy variable after that date.}
    \item $COVID$: Dummy variable with a value of 1 for the period with strict mobility restrictions due to the pandemic (from March 15, 2020 to June 21, 2020).
    \item $WAR$: Dummy variable with value 1 from February 24, 2022 onward, covering the outbreak of the war in Ukraine.
   \item $MIN$: Minimum price of the day.
    \item $MAX$: Maximum price of the day.
\end{itemize}

  The $COVID$ and $WAR$ dummy variables are considered with one lag to account for the time needed to react to new information, while the variables $W$, $S$, and $LF$ are contemporary because they represent forecasts rather than actual values. Furthermore, the maximum and minimum prices are included with one lag. Finally, $P$, lagged up to seven days for each hour, is also included as explanatory variable.
Therefore, we consider the following model specification:

\begin{equation} \label{eq:model}
\begin{aligned}
P_{t,h} =\ & \beta_1 + \beta_2 W_{t,h} + \beta_3 S_{t,h} + \beta_4 LF_{t,h} 
+ \sum_{i=1}^7 \alpha_i D_i + \sum_{i=1}^{12} \gamma_i M_i \\
& + \delta_1 DG_t + \delta_2 DP_t + \delta_3 COVID_{t-1} + \delta_4 WAR_{t-1}+\delta_5 MIN_{t-1} \\
& +\delta_6 MAX_{t-1}+ \sum_{h=1}^{24} \sum_{i=1}^7 \lambda_{i,h} P_{t-i,h} + \epsilon_{t,h} \qquad t=1,\ldots,1827,\; h=1,\ldots,24
\end{aligned}
\end{equation}

\noindent where we assume the probability distribution functions for prices described in Section \ref{specification}.

As a benchmark, we consider a naive model in which hourly electricity price forecast is the previous day's price during weekdays, and previous weekend's price during weekends. A dummy variable, $Wd$, is thus generated for weekends.
\begin{equation} \label{eq:naive}
P_{t,h} =(1-Wd) P_{t,h-1} + (Wd) P_{t-7,h} + \epsilon_{t,h} \qquad t=1,\ldots,1827,\; h=1,\ldots,24
\end{equation}
\noindent where prices are assumed to be normally distributed.

\subsection{Descriptive statistics} \label{statistics}
Table \ref{tab:stat_all} shows the main descriptive statistics of price and the variables used to model its distribution over the entire sample period.\footnote{The ADF unit root test reveals that all the variables are stationary at a 10\% significance level. Results are available from the authors upon request.} 

\begin{table}[H]
\centering \caption{Summary statistics} \label{tab:stat_all}
\begin{threeparttable}
\begin{tabular}{lrrrrrrrr}
  \hline
Variable & Mean & Median & St.Dev. & Max. & Min. & Skew. & Ex.Kurt. &  J-B \\ 
  \hline
$P$ &92.66 & 82.00 & 70.25 & 700.00 & -2.00 & 1.23$^a$ & 2.65$^a$ & 23841.91$^a$ \\ 
$S$ & 3787.55 & 628.00 & 4934.82 & 21218.95 & 0.00 & 1.23$^a$ & 0.45$^a$ & 11450.45$^a$ \\ 
$W$ & 6645.25 & 5872.88 & 3852.99 & 19753.75 & 0.00 & 0.79$^a$ & 0.03 & 4604.18$^a$ \\ 
$LF$ &  26899.57 & 26870.25 & 4339.07 & 41786.00 & 0.00 & 0.13$^a$ & -0.66$^a$ & 924.57$^a$ \\ 
   \hline
\end{tabular}
\begin{tablenotes}
\footnotesize
\item St.Dev., Max., Min., Skew., Ex.Kurt., and J-B stand for Standard Deviation, Maximum, Minimum, Skewness, Excess Kurtosis and Jarque-Bera statistic, respectively. $^a$ indicates rejection of the null hypothesis (no skewness, no excess kurtosis and no normality at the 1\% significance level).
\end{tablenotes}
\end{threeparttable}
\end{table}

All the variables exhibit skewness and excess kurtosis, indicating a non-normal distribution. The mean price is \euro92.66/MWh with maximum and minimum values of \euro700 and -\euro2/MWh, respectively. This indicates that the price limits at the start of the sample period were exceeded following the regulatory change in 2021. The standard deviation of $S$, $W$ and $LF$ is very high due to the significant differences in the figures of these variables between the hours of the day.

\begin{figure}[h!]
\caption{Electricity prices} \label{fig:prices_density}

\begin{subfigure}[t]{0.75\linewidth}
    \centering
    \includegraphics[height=6cm]{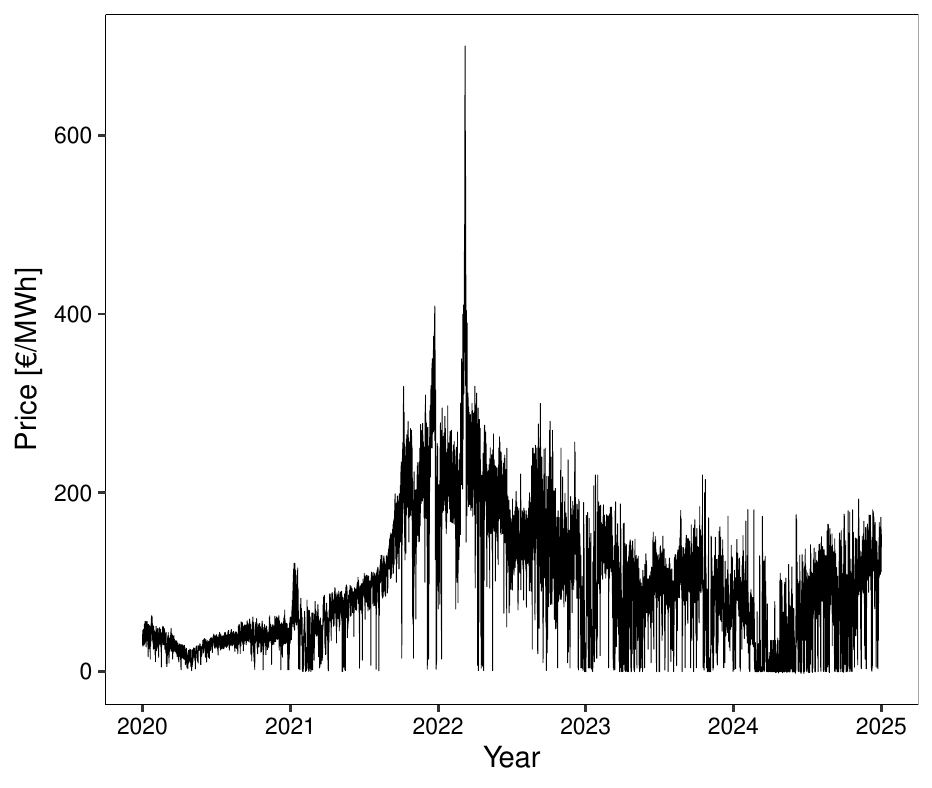}  
    \caption{Prices}
    \label{fig:prices}
\end{subfigure}%
\hspace{-6.5em}  
\begin{subfigure}[t]{0.23\linewidth}
    \centering
    \includegraphics[height=6cm]{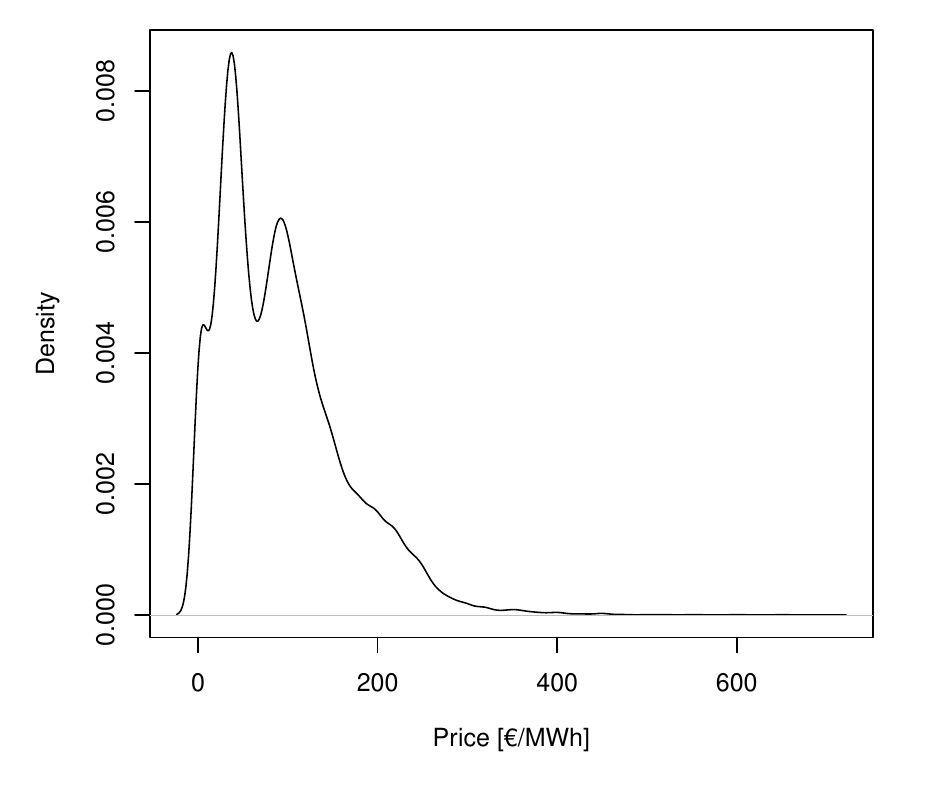}  
    \caption{Density}
    \label{fig:density}
\end{subfigure}
\end{figure}

Figure \ref{fig:prices_density} shows the hourly electricity 
prices throughout the sample period (Panel \ref{fig:prices}) and the corresponding Kernel density estimate (Panel \ref{fig:density}). The figure confirms the characteristics observed in Table \ref{tab:stat_all} such as high volatility, skewness, and kurtosis. The figure also shows that most of the density is between \euro20 and \euro125/MWh, and that the moments of the price distribution (mean, variance, skewness and kurtosis) do not remain constant during the sample period.

More detailed information can be obtained from the price by year and hour statistics. Table \ref{tab:stat_year_hour} provides the mean, standard deviation, skewness and excess kurtosis of $P$ for each year and hour in the sample. As observed before, these statistics do not remain constant throughout the years and hours. Specifically, the mean price is higher during peak hours due to higher demand, but it is also higher after the regulatory change in 2021, particularly in 2022, due to the impact of the war in Ukraine and rising gas prices. These price changes caused the highest standard deviation in 2021 and 2022. By contrast, low prices and volatility are observed in 2020 due to the drastic fall in demand as a consequence of the COVID-19 pandemic. These figures show the presence of heteroscedasticity.

Regarding skewness and kurtosis, clear differences are more noticeable year on year. Positive skewness is observed in all hours in both 2021 and 2022, reflecting the price increases in these years. High positive kurtosis was observed in 2022, revealing the presence of more frequent extreme values and non-normality. 

Therefore, the descriptive evidence supports the use of GAMLSS, as it allows not only the conditional mean but also the variance, skewness, and kurtosis of the response distribution to be modelled as functions of explanatory variables.

\begin{table}[H]
\centering \caption{Summary statistics of price by year and hour}\label{tab:stat_year_hour}
\scalebox{0.85}{
\begin{tabular}{cccccc|ccccc }
& \multicolumn{5}{c}{Mean} & \multicolumn{5}{c}{Standard Deviation} \\
Hour & 2020 & 2021 & 2022 & 2023 & 2024 & 2020 & 2021 & 2022 & 2023 & 2024 \\
\hline
1 & 33.45 & 112.96 & 176.80 & 94.59 & 71.76 & 9.17 & 73.73 & 64.97 & 35.69 & 41.22 \\
2 & 30.62 & 105.92 & 163.26 & 87.64 & 66.11 & 9.17 & 70.78 & 63.28 & 36.21 & 40.70 \\
3 & 28.79 & 100.94 & 155.74 & 83.40 & 62.67 & 9.02 & 68.38 & 62.45 & 35.98 & 39.94 \\
4 & 27.72 & 97.45 & 150.25 & 81.14 & 61.06 & 9.00 & 65.81 & 61.03 & 35.68 & 40.13 \\
5 & 27.36 & 96.14 & 149.13 & 79.57 & 59.61 & 9.09 & 65.08 & 60.83 & 35.61 & 39.93 \\
6 & 28.44 & 99.52 & 154.17 & 82.53 & 61.58 & 9.37 & 67.54 & 62.65 & 35.91 & 39.86 \\
7 & 31.48 & 107.43 & 167.00 & 91.18 & 68.29 & 10.42 & 71.49 & 68.11 & 36.21 & 41.01 \\
8 & 34.50 & 116.80 & 182.77 & 101.63 & 78.88 & 11.88 & 76.63 & 71.37 & 37.23 & 42.77 \\
9 & 36.05 & 121.33 & 186.73 & 104.27 & 78.22 & 12.42 & 77.48 & 71.74 & 38.96 & 45.82 \\
10 & 36.55 & 120.13 & 178.72 & 92.70 & 65.24 & 12.05 & 77.18 & 69.73 & 39.10 & 45.37 \\
11 & 35.67 & 115.13 & 165.15 & 78.95 & 51.98 & 11.82 & 74.56 & 68.08 & 39.23 & 42.89 \\
12 & 35.01 & 110.41 & 155.02 & 71.14 & 43.64 & 11.52 & 72.24 & 66.12 & 39.11 & 39.90 \\
13 & 34.86 & 108.62 & 152.33 & 67.87 & 40.65 & 11.19 & 71.87 & 65.26 & 38.74 & 38.92 \\
14 & 34.52 & 106.85 & 149.64 & 65.24 & 38.92 & 11.03 & 71.47 & 63.44 & 37.92 & 38.61 \\
15 & 33.37 & 103.90 & 144.06 & 62.16 & 37.85 & 10.95 & 71.18 & 63.98 & 37.91 & 39.31 \\
16 & 31.93 & 100.25 & 140.21 & 60.33 & 37.73 & 11.58 & 72.06 & 63.88 & 38.69 & 40.90 \\
17 & 32.06 & 102.10 & 143.42 & 64.66 & 42.46 & 12.15 & 74.92 & 65.77 & 39.88 & 44.61 \\
18 & 33.92 & 110.75 & 155.14 & 76.26 & 51.90 & 12.74 & 79.46 & 71.03 & 42.52 & 48.56 \\
19 & 36.42 & 120.13 & 171.63 & 89.89 & 64.56 & 12.84 & 82.93 & 73.24 & 43.55 & 50.97 \\
20 & 38.96 & 128.63 & 190.65 & 107.57 & 79.65 & 12.13 & 82.91 & 74.01 & 42.30 & 49.43 \\
21 & 40.37 & 132.47 & 207.36 & 119.50 & 91.51 & 11.31 & 82.57 & 69.09 & 36.71 & 46.62 \\
22 & 40.04 & 130.16 & 208.80 & 118.31 & 94.81 & 9.81 & 78.77 & 67.39 & 33.16 & 43.30 \\
23 & 38.04 & 123.27 & 194.93 & 109.72 & 86.30 & 8.91 & 74.60 & 65.97 & 32.86 & 41.43 \\
24 & 34.90 & 115.03 & 177.99 & 100.12 & 77.39 & 8.60 & 70.69 & 64.94 & 33.56 & 40.76 \\ 
Total & 33.96 & 111.93 & 167.54 & 87.10 & 63.03 & 11.41 & 74.72 & 69.44 & 41.35 & 45.93 \\
\hline
& \multicolumn{5}{c|}{Skewness} & \multicolumn{5}{c}{Excess Kurtosis} \\
Hour & 2020 & 2021 & 2022 & 2023 & 2024 & 2020 & 2021 & 2022 & 2023 & 2024 \\
\hline
1 & -0.55 & 1.02 & 0.62 & -0.86 & -0.26 & 0.08 & 0.61 & 3.61 & 0.11 & -1.05 \\
2 & -0.48 & 0.99 & 0.50 & -0.78 & -0.18 & 0.09 & 0.56 & 2.92 & -0.09 & -1.17 \\
3 & -0.49 & 0.96 & 0.54 & -0.73 & -0.13 & 0.13 & 0.52 & 2.89 & -0.19 & -1.18 \\
4 & -0.51 & 0.86 & 0.50 & -0.75 & -0.09 & 0.11 & 0.28 & 2.64 & -0.22 & -1.23 \\
5 & -0.57 & 0.86 & 0.53 & -0.70 & -0.07 & 0.06 & 0.34 & 2.86 & -0.26 & -1.26 \\
6 & -0.59 & 0.94 & 0.60 & -0.71 & -0.09 & 0.05 & 0.65 & 3.21 & -0.19 & -1.21 \\
7 & -0.39 & 0.97 & 1.09 & -0.81 & -0.16 & -0.09 & 0.66 & 5.45 & 0.12 & -1.08 \\
8 & -0.32 & 1.00 & 1.28 & -0.92 & -0.31 & -0.40 & 0.48 & 6.38 & 0.57 & -0.92 \\
9 & -0.26 & 1.03 & 1.39 & -0.76 & -0.22 & -0.51 & 0.48 & 6.87 & 0.39 & -1.01 \\
10 & -0.31 & 1.06 & 1.26 & -0.58 & 0.03 & -0.46 & 0.58 & 5.41 & -0.19 & -1.11 \\
11 & -0.33 & 1.11 & 1.20 & -0.32 & 0.30 & -0.42 & 0.93 & 4.94 & -0.63 & -1.09 \\
12 & -0.35 & 1.14 & 1.09 & -0.23 & 0.45 & -0.38 & 1.21 & 4.51 & -0.73 & -0.95 \\
13 & -0.38 & 1.13 & 0.97 & -0.21 & 0.51 & -0.38 & 1.21 & 3.86 & -0.84 & -0.91 \\
14 & -0.43 & 1.14 & 0.80 & -0.21 & 0.58 & -0.28 & 1.27 & 3.01 & -0.90 & -0.88 \\
15 & -0.41 & 1.13 & 0.75 & -0.14 & 0.66 & -0.20 & 1.34 & 2.66 & -0.93 & -0.79 \\
16 & -0.35 & 1.21 & 0.67 & -0.09 & 0.75 & -0.31 & 1.55 & 2.42 & -1.02 & -0.64 \\
17 & -0.30 & 1.18 & 0.97 & -0.14 & 0.74 & -0.48 & 1.27 & 4.00 & -0.92 & -0.61 \\
18 & -0.26 & 1.05 & 1.47 & -0.23 & 0.56 & -0.60 & 0.62 & 6.69 & -0.75 & -0.88 \\
19 & -0.18 & 1.04 & 1.63 & -0.32 & 0.34 & -0.67 & 0.33 & 7.37 & -0.53 & -0.93 \\
20 & -0.39 & 1.01 & 1.86 & -0.22 & -0.10 & -0.55 & 0.10 & 9.14 & 0.05 & -1.05 \\
21 & -0.55 & 1.01 & 1.47 & -0.46 & -0.26 & -0.41 & 0.05 & 7.23 & 0.58 & -0.79 \\
22 & -0.48 & 0.96 & 1.28 & -1.00 & -0.44 & -0.28 & 0.09 & 7.38 & 1.20 & -0.64 \\
23 & -0.54 & 0.96 & 1.27 & -1.09 & -0.49 & -0.10 & 0.31 & 8.15 & 1.09 & -0.70 \\
24 & -0.45 & 0.94 & 1.10 & -0.94 & -0.36 & -0.23 & 0.38 & 6.41 & 0.49 & -0.89 \\ 
Total & -0.24 & 1.06 & 1.03 & -0.41 & 0.12 & -0.28 & 0.75 & 4.98 & -0.38 & -1.09 \\
\hline
\end{tabular}
}
\end{table}

\section{Results} \label{results}

This section presents the results of the error measures calculated using price forecasts generated from a two-year rolling window, as described in Section \ref{forecast}. Table \ref{errors-mean} reports the average error measures across hours based on the regression model (\ref{eq:model}) and the probability distribution functions described in Section \ref{specification}. The table also includes the results for the naive model (Equation (\ref{eq:naive})) assuming a normal probability distribution\footnote{Results for each hour are reported in Appendix \ref{app:error-hour}.}.

In terms of point forecasts, the SHASH distribution yields the highest price forecasting accuracy, closely followed by the JSU(4) distribution. In contrast, the naive model performs worst overall, as it is outperformed by model (\ref{eq:model}) under all considered probability distributions (NO, JSU, and SHASH). These results highlight the advantage of modelling all four parameters of the price distribution for forecasting purposes. 

When assuming the JSU distribution, forecast accuracy improves as additional  distributional parameters are allowed to vary with the explanatory variables. In particular, the MAE decreases from the JSU(2) specification, where the shape parameters are constants, to the JSU(4) specification, in which all four parameters are modelled as functions of the explanatory variables.

\begin{table}[H]
\centering
\caption{Forecast error measures by distribution} \label{errors-mean}
\hspace{-3cm}
\begin{threeparttable}
\begin{tabular}{rllllll}
  \hline
 & Naive & NO & JSU(2) & JSU(3) & JSU(4) & SHASH(2) \\ 
  \hline
  MAE & $\cellcolor[rgb]{1,0.5,0.5} \underset{(0.17)}{25.29}$ & $\cellcolor[rgb]{0.517,1,0.5} \underset{(0.12)}{19.44}$ & $\cellcolor[rgb]{0.929,1,0.5} \underset{(2.43)}{21.89}$ & $\cellcolor[rgb]{0.561,1,0.5} \underset{(0.30)}{19.70}$ & $\cellcolor[rgb]{0.508,1,0.5} \underset{(0.12)}{19.39}$ & $\cellcolor[rgb]{0.5,1,0.5} \underset{(0.12)}{19.34}$ \\ 
   \hline
    $PB_{\tau=0.99}$ & $\cellcolor[rgb]{1,0.5,0.5} \underset{(0.04)}{1.76}$ & $\cellcolor[rgb]{1,0.952,0.5} \underset{(0.04)}{1.55}$ & $\cellcolor[rgb]{0.686,1,0.5} \underset{(0.04)}{1.38}$ & $\cellcolor[rgb]{1,0.775,0.5} \underset{(0.27)}{1.63}$ & $\cellcolor[rgb]{0.5,1,0.5} \underset{(0.03)}{1.29}$ & $\cellcolor[rgb]{0.772,1,0.5} \underset{(0.03)}{1.42}$ \\ 
   \hline
    $PB_{\tau=0.95}$ & $\cellcolor[rgb]{1,0.5,0.5} \underset{(0.06)}{4.84}$ & $\cellcolor[rgb]{0.5,1,0.5} \underset{(0.05)}{3.41}$ & $\cellcolor[rgb]{0.805,1,0.5} \underset{(0.13)}{3.85}$ & $\cellcolor[rgb]{0.906,1,0.5} \underset{(0.26)}{3.99}$ & $\cellcolor[rgb]{0.501,1,0.5} \underset{(0.04)}{3.41}$ & $\cellcolor[rgb]{0.922,1,0.5} \underset{(0.05)}{4.01}$ \\ 
   \hline
   $PB_{\tau=0.05}$ & $\cellcolor[rgb]{0.978,1,0.5} \underset{(0.06)}{4.68}$ & $\cellcolor[rgb]{0.66,1,0.5} \underset{(0.05)}{3.92}$ & $\cellcolor[rgb]{1,0.5,0.5} \underset{(2.30)}{5.95}$ & $\cellcolor[rgb]{0.56,1,0.5} \underset{(0.05)}{3.67}$ & $\cellcolor[rgb]{0.5,1,0.5} \underset{(0.05)}{3.53}$ & $\cellcolor[rgb]{0.638,1,0.5} \underset{(0.05)}{3.86}$ \\ 
   \hline
   $PB_{\tau=0.01}$  & $\cellcolor[rgb]{0.723,1,0.5} \underset{(0.04)}{1.66}$ & $\cellcolor[rgb]{0.809,1,0.5} \underset{(0.04)}{1.87}$ & $\cellcolor[rgb]{1,0.5,0.5} \underset{(2.35)}{3.53}$ & $\cellcolor[rgb]{0.53,1,0.5} \underset{(0.02)}{1.20}$ & $\cellcolor[rgb]{0.5,1,0.5} \underset{(0.02)}{1.13}$ & $\cellcolor[rgb]{0.548,1,0.5} \underset{(0.02)}{1.24}$ \\ 
   \hline
\end{tabular}
\begin{tablenotes}
\footnotesize
\item Note: NO, JSU, and SHASH denote the Normal, Johnson's SU, and Sinh–Arcsinh distributions, respectively. The number in parentheses next to each distribution name indicates how many distribution moments are modelled as non-constant, i.e., as functions of the explanatory variables. For example, JSU(2) means that the first two moments are non-constant. Parentheses beneath the error measures contain the corresponding standard deviations. A heat map is used to indicate lower (green) and higher (red) forecast errors.
\end{tablenotes}
\end{threeparttable}
\end{table}

Regarding the probabilistic forecasting performance, the average PB across all hours is reported for the quantiles: $\tau = 0.99, 0.95, 0.05,$ and $ 0.01$, thereby assessing forecasting accuracy in both the upper and lower tails of the price distribution. 

For the upper tail, the JSU(4) distribution provides the most accurate forecasts at the 0.99 quantile, followed by the JSU(2) distribution, which provides accurate forecasts of the most extreme high prices, despite its poor performance when forecasting central price values (MAE = 21.89). Similar results hold for the 0.95 quantile, except that the NO distribution, together with the JSU(4) distribution, is the best at forecasting high prices. In contrast, the naive model performs worst for both upper-tail quantiles, indicating limited ability to predict positive price spikes.

For the lower tail of the distribution (the 0.05 and 0.01 quantiles), the JSU(4) specification again outperforms all the other distributions and the naive model. The JSU(3) and SHASH(2) distributions also produce accurate forecasts, whereas JSU(2) distribution performs very poorly for extreme low prices.

Overall, the results suggest that model (\ref{eq:model}) under the JSU distribution, where all four distributional parameters are modelled as functions of the explanatory variables, provides the best probabilistic forecasting performance, particularly in both tails of the distribution. Moreover, it is the second-best distribution in terms of point forecast accuracy, with a MAE value very similar to that of the SHASH distribution. These findings emphasise the benefits of using the GAMLSS framework to model distribution and forecast prices.

Finally, as mentioned in Section \ref{method}, we use the DM test with the Harvey-Leybourne-Newbold correction to formally assess differences in forecasting performance across all pairs of competing models under each loss criterion. The corresponding p-values for the calculations pooled over 24 hours are reported in Tables \ref{tab:dm_test_mae}--\ref{tab:dm_test_1}. 

For the point forecasts based on the MAE loss function, the naive benchmark performs significantly worse than model (\ref{eq:model}) under all  distributional assumptions except JSU(2). Although SHASH(2) achieves a slightly lower MAE than JSU(4), the DM test indicates that the difference is not statistically significant. More generally, no significant differences are found among the JSU(2), JSU(3), JSU(4), and SHASH(2) specifications, suggesting that their point forecasting performances are broadly comparable. However, both JSU(4) and SHASH(2) distributions significantly outperform the NO specification, providing evidence that greater distributional flexibility improves point forecast accuracy.

The results provide stronger support for the JSU(4) specification, for probabilistic forecasts using the $\tau$-level PB loss function. In the upper tail, JSU(4) significantly outperforms the NO, JSU(2), and SHASH(2) distributions at the 99\% quantile, while at the 95\% quantile it significantly outperforms JSU(2), JSU(3), and SHASH(2). In the lower tail, JSU(4) significantly outperforms both JSU(3) and SHASH(2) at the 5\% and 1\% quantiles. Although some pairwise differences involving JSU(2) are not statistically significant, the average pinball losses reported in Table \ref{errors-mean} consistently favour JSU(4).

The DM test also confirms that the naive benchmark generally performs worse than the GAMLSS specifications, particularly for forecasting the upper tail of the price distribution. However, some exceptions arise in the lower tail, where the naive model is not significantly different from JSU(2) and even outperforms the NO specification at the 1\% quantile.

Overall, the DM test results reinforce the conclusions obtained from the forecast error measures. While differences in point forecasting accuracy among the best-performing specifications are relatively small, the JSU(4) model provides the most robust probabilistic forecasting performance across both tails of the distribution. These findings highlight the benefits of modelling all four distributional parameters within the GAMLSS framework when forecasting extreme electricity prices.


Looking at the error measures and DM tests by hour (see Tables \ref{tab:mae_hour}--\ref{tab:PB1_hour} in Appendix \ref{app:error-hour}), the MAE results reveal substantial intraday variation in forecasting performance across the specifications. Although SHASH(2) achieves the lowest average MAE across all hours, no single specification consistently outperforms the other at every hour of the day. The NO specification performs particularly well during the morning hours, whereas SHASH(2) tends to provide the most accurate forecasts during midday and early-night periods. The JSU specifications become relatively more competitive during the evening peak hours. Overall, the hourly analysis supports the conclusions drawn from the aggregate MAE results, namely that the SHASH(2) and JSU(4) provide the best point forecasting performance. At the same time, it highlights that the relative advantages of the alternative distributions vary across the daily price cycle.

By contrast, when probabilistic forecast is considered, the JSU(4) specification provides the lowest pinball losses across most hours and quantiles. Moreover, the DM tests indicate that the small MAE advantage of SHASH(2) is not statistically significant, whereas the improvements achieved by JSU(4) in forecasting extreme quantiles are statistically significant in a large number of hours. These results indicate that modelling all four distributional parameters as functions of explanatory variables yields substantial gains in characterising tail behaviour and forecasting extreme price movements. Consequently, JSU(4) emerges as the preferred overall specification when both point and probabilistic forecasting performance are taken into account.


\begin{table}[H]
\centering
\caption{DM test p-values for MAE loss criterion} \label{tab:dm_test_mae}
\begin{threeparttable}
\begin{tabular}{lccccc}
  \toprule
 & NO & JSU(2) & JSU(3) & JSU(4) & SHASH(2) \\ 
  \midrule
Naive   &   0.000$^{a}$ $\uparrow$ & 0.162 $\uparrow$ & 0.000$^{a}$ $\uparrow$ & 0.000$^{a}$ $\uparrow$ & 0.000$^{a}$ $\uparrow$ \\ 
  NO    &   —          & 0.313 $\downarrow$ & 0.340 $\downarrow$ & 0.094$^{c}$ $\uparrow$ & 0.017$^{b}$ $\uparrow$ \\ 
  JSU(2) &  — &  — & 0.371 $\uparrow$ & 0.299 $\uparrow$ & 0.293 $\uparrow$ \\ 
  JSU(3) &  — &  — &  — & 0.223 $\uparrow$ & 0.185 $\uparrow$ \\ 
  JSU(4) &  — &  — &  — &  — & 0.372 $\uparrow$ \\ 
   \bottomrule
\end{tabular}
\begin{tablenotes}
\item Note: DM test with Harvey-Leybourne-Newbold correction for MAE loss criterion. $H_0$: equal predictive accuracy between row and column model. $^{a}$, $^{b}$ and $^{c}$ show rejection of $H_0$ at 1\%, 5\% and 10\%. $\downarrow$ ($\uparrow$) indicates the row model has lower (higher) mean loss.
\end{tablenotes}
\end{threeparttable}
\end{table}

\begin{table}[H]
\centering
\caption{DM test p-values for 99\% Pinball loss criterion } 
\label{tab:dm_test_99}
\begin{threeparttable}
\begin{tabular}{lccccc}
  \toprule
&  NO & JSU(2) & JSU(3) & JSU(4) & SHASH(2) \\  \midrule
Naive    & 0.000$^{a}$ $\uparrow$ & 0.000$^{a}$ $\uparrow$ & 0.633 $\uparrow$ & 0.000$^{a}$ $\uparrow$ & 0.000$^{a}$ $\uparrow$ \\ 
  NO     & — & 0.000$^{a}$ $\uparrow$ & 0.756 $\downarrow$ & 0.000$^{a}$ $\uparrow$ & 0.000$^{a}$ $\uparrow$ \\ 
  JSU(2) & — & — & 0.344 $\downarrow$ & 0.010$^{b}$ $\uparrow$ & 0.196 $\downarrow$ \\ 
  JSU(3) & — & — & — & 0.202 $\uparrow$ & 0.425 $\uparrow$ \\ 
  JSU(4) & — & — & — & — & 0.000$^{a}$ $\downarrow$ \\ 
   \bottomrule
\end{tabular}
\begin{tablenotes}
\item Note: DM test with Harvey-Leybourne-Newbold correction for 99\% Pinball loss criterion. $H_0$: equal predictive accuracy between row and column model. $^{a}$ and $^{b}$ show rejection of $H_0$ at 1\% and 5\%. $\downarrow$ ($\uparrow$) indicates the row model has lower (higher) mean loss.
\end{tablenotes}
\end{threeparttable}
\end{table}

\begin{table}[H]
\centering
\caption{DM test p-values for 95\% Pinball loss criterion} 
\label{tab:dm_test_95}
\begin{threeparttable}
\begin{tabular}{lccccc}
  \toprule
 & NO & JSU(2) & JSU(3) & JSU(4) & SHASH(2) \\ 
  \midrule
Naive   & 0.000$^{a}$ $\uparrow$ & 0.000$^{a}$ $\uparrow$ & 0.001$^{a}$ $\uparrow$ & 0.000$^{a}$ $\uparrow$ & 0.000$^{a}$ $\uparrow$ \\ 
  NO    & — & 0.000$^{a}$ $\downarrow$ & 0.026$^{b}$ $\downarrow$ & 0.901 $\uparrow$ & 0.000$^{a}$ $\downarrow$ \\ 
  JSU(2) & — & — & 0.614 $\downarrow$ & 0.001$^{a}$ $\uparrow$ & 0.183 $\downarrow$ \\ 
  JSU(3) & — & — & — & 0.027$^{b}$ $\uparrow$ & 0.931 $\downarrow$ \\ 
  JSU(4) & — & — & — & — & 0.000$^{a}$ $\downarrow$ \\ 
   \bottomrule
\end{tabular}
\begin{tablenotes}
\item Note: DM test with Harvey-Leybourne-Newbold correction for 95\% Pinball loss criterion. $H_0$: equal predictive accuracy between row and column model. $^{a}$ and $^{b}$ show rejection of $H_0$ at 1\% and 5\%.  $\downarrow$ ($\uparrow$) indicates the row model has lower (higher) mean loss.
\end{tablenotes}
\end{threeparttable}
\end{table}

\begin{table}[H]
\centering
\caption{DM test p-values for 5\% Pinball loss criterion} 
\label{tab:dm_test_5}
\begin{threeparttable}
\begin{tabular}{lccccc}
  \toprule
 & NO & JSU(2) & JSU(3) & JSU(4) & SHASH(2) \\
  \midrule
Naive   & 0.000$^{a}$ $\uparrow$ & 0.584 $\downarrow$ & 0.000$^{a}$ $\uparrow$ & 0.000$^{a}$ $\uparrow$ & 0.000$^{a}$ $\uparrow$ \\ 
  NO     & — & 0.378 $\downarrow$ & 0.000$^{a}$ $\uparrow$ & 0.000$^{a}$ $\uparrow$ & 0.262 $\uparrow$ \\ 
  JSU(2) & — & — & 0.324 $\uparrow$ & 0.295 $\uparrow$ & 0.366 $\uparrow$ \\ 
  JSU(3) & — & — & — & 0.000$^{a}$ $\uparrow$ & 0.000$^{a}$ $\downarrow$ \\ 
  JSU(4) & — & — & — & — & 0.000$^{a}$ $\downarrow$ \\ 
   \bottomrule
\end{tabular}
\begin{tablenotes}
\item Note: DM test with Harvey-Leybourne-Newbold correction for 5\% Pinball loss criterion. $H_0$: equal predictive accuracy between row and column model. $^{a}$ shows rejection of $H_0$ at 1\%. $\downarrow$ ($\uparrow$) indicates the row model has lower (higher) mean loss.T
\end{tablenotes}
\end{threeparttable}
\end{table}

\begin{table}[H]
\centering
\caption{DM test p-values for 1\% Pinball loss criterion} 
\label{tab:dm_test_1}
\begin{threeparttable}
\begin{tabular}{lccccc}
  \toprule
 & NO & JSU(2) & JSU(3) & JSU(4) & SHASH(2) \\ 
  \midrule
Naive  & 0.000$^{a}$ $\downarrow$ & 0.428 $\downarrow$ & 0.000$^{a}$ $\uparrow$ & 0.000$^{a}$ $\uparrow$ & 0.000$^{a}$ $\uparrow$ \\ 
  NO   & — & 0.481 $\downarrow$ & 0.000$^{a}$ $\uparrow$ & 0.000$^{a}$ $\uparrow$ & 0.000$^{a}$ $\uparrow$ \\ 
  JSU(2) & — & — & 0.322 $\uparrow$ & 0.308 $\uparrow$ & 0.331 $\uparrow$ \\ 
  JSU(3) & — & — & — & 0.001$^{a}$ $\uparrow$ & 0.020$^{b}$ $\downarrow$ \\ 
  JSU(4) & — & — & — & — & 0.000$^{a}$ $\downarrow$ \\ 
   \bottomrule
\end{tabular}
\begin{tablenotes}
\item Note: DM test with Harvey-Leybourne-Newbold correction for 1\% Pinball loss criterion. $H_0$: equal predictive accuracy between row and column model. $^{a}$ and $^{b}$ show rejection of $H_0$ at 1\% and 5\%. $\downarrow$ ($\uparrow$) indicates the row model has lower (higher) mean loss.
\end{tablenotes}
\end{threeparttable}
\end{table}


\section{Conclusions} \label{conclusions}

This paper applies the GAMLSS framework to forecast electricity prices in the Spanish day-ahead market. This market constitutes a particularly interesting case study due to the substantial changes experienced during the period analysed (2020-2024), including the COVID-19 lockdown, revisions to market price limits, the war in Ukraine, and the implementation of the Iberian exception. These events triggered pronounced changes not only in the level and volatility of prices, but also in their asymmetry and kurtosis through the sample period. This creates a suitable environment for evaluating forecasting approaches capable of capturing time-varying distributional characteristics.

The GAMLSS approach allows the entire conditional distribution of prices to be modelled by linking its parameters to market fundamentals. Specifically, we examine alternative distributional specifications in which the location, scale, skewness, and kurtosis parameters are allowed to vary dynamically with explanatory variables related to electricity market conditions. To this end, several GAMLSS specifications based on the Normal, Johnson's SU (JSU), and Sinh–Arcsinh (SHASH) distributions are estimated using market fundamentals, including demand forecasts, renewable generation forecasts, seasonal effects, and variables capturing major regulatory and geopolitical events. Alternative specifications were considered according to the number of distributional parameters allowed to vary with explanatory variables.

The empirical results provide strong evidence that modelling higher-order moments of the electricity price distribution improves forecasting performance. For point forecasts, the SHASH(2) and JSU(4) distributions achieve the lowest forecasting errors and substantially outperform both the naive benchmark and the standard normal specification. The hourly analysis further reveals that forecasting performance varies considerably across the daily price cycle. While SHASH(2) achieves the lowest average MAE, no single specification dominates at all hours. The NO specification performs particularly well during the morning hours, SHASH(2) tends to provide the most accurate forecasts during midday and early-night periods, and the JSU specifications become relatively more competitive during the evening peak hours. These results suggest that the benefits of alternative distributional assumptions depend on prevailing market conditions throughout the day.

The benefits of flexible distributional modelling are even more pronounced in probabilistic forecasting. The JSU(4) specification, in which all four distributional parameters depend on market fundamentals, consistently yields the most accurate forecasts in both the upper and lower tails of the distribution. The Diebold--Mariano tests show that the small MAE advantage of SHASH(2) over JSU(4) is generally not statistically significant, whereas the gains achieved by JSU(4) in forecasting extreme quantiles are statistically significant in a large number of cases. These findings indicate that allowing all four distributional parameters to evolve with market fundamentals substantially improves the characterization of tail behaviour and the prediction of extreme price movements.

These findings highlight the importance of explicitly modelling asymmetry and tail behaviour in electricity markets. Point forecasts alone provide limited insight in environments characterised by frequent price spikes, negative prices, and changing volatility. By allowing the conditional skewness and kurtosis to evolve over time in response to market conditions, the GAMLSS framework offers a richer representation of market uncertainty and enhances the forecasting of rare but economically important events.

Overall, the evidence shows that flexible distributional modelling constitutes a valuable approach for electricity price forecasting in the Spanish market. By capturing the time-varying behaviour of the entire conditional price distribution, GAMLSS models improve the accuracy of forecasts for both typical and extreme market conditions, thereby offering useful information for decision-making in increasingly volatile and renewable-intensive electricity systems.

The results also have practical implications for market participants and policy makers. For financial market participants, the superior ability of JSU(4) to capture asymmetry and tail behaviour implies more accurate estimates of downside and upside risk, leading to improved hedging design decisions. For regulators, the findings are particularly relevant in the context of the ongoing reform of the European electricity market, where strengthening long-term contracting mechanisms and risk-sharing instruments requires reliable assessments of future price distributions rather than forecasts of expected prices alone. In both cases, as renewable penetration increases and market outcomes become more volatile, accurate modelling of the entire predictive distribution becomes increasingly important, since uncertainty is a key variable affecting market efficiency and financial stability. Taken together, the results suggest that JSU(4) constitutes the preferred specification when both point and probabilistic forecasting performance are considered, demonstrating the value of flexible distributional modelling for decision-making in increasingly volatile and renewable-intensive electricity systems.

\section*{Funding}

This work was supported by MICIU/AEI/10.13039/501100011033 and by ERDF, EU, through grant PID2022-139458NB-I00, and by the Basque Government through research grant IT1461-22.


\section*{Declaration of generative AI and AI-assisted technologies in the manuscript preparation process}

During the preparation of this work the authors used ChatGPT (OpenAI, GPT-5.5) in order to assist with language editing, grammar correction, and improvement of text clarity and readability. After using this tool/service, the authors reviewed and edited the content as needed and take full responsibility for the content of the published article.

\break

\bibliographystyle{unsrtnat}  
\bibliography{bibliography}

@article{Aggarwal2009,
  author    = {S. K. Aggarwal and L. M. Saini and A. Kumar},
  title     = {Electricity Price Forecasting in Deregulated Markets: A Review and Evaluation},
  journal   = {International Journal of Electrical Power and Energy Systems},
  volume    = {31},
  number    = {1},
  pages     = {13--22},
  year      = {2009},
  doi       = {10.1016/j.ijepes.2008.09.003}
}

@article{Cifter2013,
  author    = {Cifter, A.},
  title     = {Forecasting electricity price volatility with the Markov-switching GARCH model: Evidence from the Nordic electric power market},
  journal   = {Electric Power Systems Research},
  volume    = {102},
  pages     = {61--67},
  year      = {2013},
  doi       = {10.1016/j.epsr.2013.04.007}
}

@article{Contreras2003,
  author    = {J. Contreras and R. Espínola and F. J. Nogales and A. J. Conejo},
  title     = {{ARIMA Models to Predict Next-Day Electricity Prices}},
  journal   = {IEEE Transactions on Power Systems},
  volume    = {18},
  number    = {3},
  pages     = {1014--1020},
  year      = {2003},
  doi       = {10.1109/TPWRS.2002.804943}
}

@misc{EC_EMD_2023,
  author       = {{European Commission}},
  title        = {{Proposal for a Regulation of the European Parliament and of the Council amending Regulations (EU) 2019/943 and (EU) 2019/942 as well as Directives (EU) 2018/2001 and (EU) 2019/944 to improve the Union's electricity market design}},
  howpublished = {COM(2023) 148 final},
  year         = {2023},
  note         = {2023/0077 (COD)},
  url          = {https://eur-lex.europa.eu/legal-content/EN/TXT/?uri=COM:2023:148:FIN}
}

@book{Hastie1990,
  title     = {Generalized Additive Models},
  author    = {Hastie, T. J. and Tibshirani, R. J.},
  year      = {1990},
  publisher = {Chapman and Hall},
  address   = {London}
}

@article{HirschZiel2024,
  author    = {Hirsch, S. and  Ziel, F.},
  title     = {{Simulation-based Forecasting for Intraday Power Markets: Modelling Fundamental Drivers for Location, Scale and Shape of the Price Distribution}},
  journal   = {The Energy Journal},
  volume={45},
  number={3},
  pages={87--124},
  year      = {2024},
  doi={10.5547/01956574.45.3.shir}
}

@article{Hong2016,
  title={{Probabilistic Energy Forecasting: Global Energy Forecasting
          Competition 2014 and Beyond}},
  author={Hong, T. and Pinson, P. and Fan, S. and Zareipour, H.
          and Troccoli, A. and Hyndman, R.J.},
  journal={International Journal of Forecasting},
  volume={32},
  number={3},
  pages={896--913},
  year={2016},
  doi={10.1016/j.ijforecast.2016.02.001}
}

@article{Johnson1949,
  author  = {Johnson, N. L.},
  title   = {{Systems of Frequency Curves Generated by Methods of Translation}},
  journal = {Biometrika},
  volume  = {36},
  number  = {1--2},
  pages   = {149--176},
  year    = {1949},
  doi     = {10.1093/biomet/36.1-2.149}
}

@article{Jones2009,
  author  = {Jones, M. C. and Pewsey, A.},
  title   = {Sinh-arcsinh distributions},
  journal = {Biometrika},
  volume  = {96},
  number  = {4},
  pages   = {761--780},
  year    = {2009},
  doi     = {10.1093/biomet/asp053}
}

@article{Lago2018,
  author    = {Lago, J. and De Ridder, F. and De Schutter, B.},
  title     = {{Forecasting spot electricity prices: Deep learning approaches and empirical comparison of traditional algorithms}},
  journal   = {Applied Energy},
  volume    = {221},
  pages     = {386--405},
  year      = {2018},
  doi       = {10.1016/j.apenergy.2018.02.069}
}

@article{Lago2021,
  title={{Forecasting Day-Ahead Electricity Prices: A Review of
          State-of-the-Art Algorithms, Best Practices and an
          Open-Access Benchmark}},
  author={Lago, J. and Marcjasz, G. and {De Schutter}, B. and Weron, R.},
  journal={Applied Energy},
  volume={293},
  pages={116983},
  year={2021},
  doi={10.1016/j.apenergy.2021.116983}
}

@article{MuniainZiel2020,
  title={{Probabilistic Forecasting in Day-Ahead Electricity Markets:
          Simulating Peak and Off-Peak Prices}},
  author={Muniain, P. and Ziel, F.},
  journal={International Journal of Forecasting},
  volume={36},
  number={4},
  pages={1193--1210},
  year={2020},
  doi={10.1016/j.ijforecast.2019.11.006}
}

@article{NarajewskiZiel2020,
  author  = {Narajewski, M. and Ziel, F.},
  title   = {Ensemble Forecasting for Intraday Electricity Prices: Simulating Trajectories},
  journal = {Applied Energy},
  volume  = {279},
  pages   = {115801},
  year    = {2020},
  doi     = {10.1016/j.apenergy.2020.115801}
}

@article{Nelder1972,
  title={{Generalized Linear Models}},
  author={Nelder, J.A. and Wedderburn, R.W.M.},
  journal={Journal of the Royal Statistical Society. Series A (General)},
  volume={135},
  number={3},
  pages={370--384},
  year={1972},
 url={https://www.jstor.org/stable/2344614}
  }

@article{Nowotarski2018,
  title={{Recent Advances in Electricity Price Forecasting:
          A Review of Probabilistic Forecasting}},
  author={Nowotarski, J. and Weron, R.},
  journal={Renewable and Sustainable Energy Reviews},
  volume={81},
  pages={1548--1568},
  year={2018},
  doi={10.1016/j.rser.2017.05.234}
}

@article{Pai2005,
  author    = {Pai, P.-F. and Hong, W.-C.},
  title     = {Support vector machines with simulated annealing algorithms in electricity load forecasting},
  journal   = {Energy Conversion and Management},
  volume    = {46},
  number    = {17},
  pages     = {2669--2688},
  year      = {2005},
  doi       = {10.1016/j.enconman.2005.02.004}
}

@article{Rigby2005,
  title={{Generalized Additive Models for Location, Scale and Shape}},
  author={Rigby, R.A. and Stasinopoulos, D.M.},
  journal={Journal of the Royal Statistical Society: Series C (Applied Statistics)},
  volume={54},
  number={3},
  pages={507--554},
  year={2005},
 doi={10.1111/j.1467-9876.2005.00510.x}
  }

@book{Rigby2020,
  author    = {R. A. Rigby and M. D. Stasinopoulos and G. Z. Heller and F. De Bastiani},
  title     = {Distributions for Modeling Location, Scale, and Shape. Using GAMLSS in R},
  year      = {2020},
  publisher = {Chapman \& Hall/CRC},
  address   = {Boca Raton, FL, USA},
  isbn      = {9780367278847}
}

@article{Serinaldi2011,
  author    = {Serinaldi, F.},
  title     = {{Distributional modeling and short-term forecasting of electricity prices by Generalized Additive Models for Location, Scale and Shape}},
  journal   = {Energy Economics},
  volume    = {33},
  number    = {6},
  pages     = {1216--1226},
  year      = {2011},
  doi       = {10.1016/j.eneco.2011.05.001}
  }

@article{Tan2010,
  author    = {Z. Tan and J. Zhang and J. Wang and J. Xu},
  title     = {{Day-ahead electricity price forecasting using wavelet transform combined with ARIMA and GARCH models}},
  journal   = {Applied Energy},
  volume    = {87},
  number    = {11},
  pages     = {3606--3610},
  year      = {2010},
  doi       = {10.1016/j.apenergy.2010.05.012}
}

@article{Tibshirani1996,
  title={{Regression Shrinkage and Selection via the Lasso}},
  author={Tibshirani, R.},
  journal={Journal of the Royal Statistical Society: Series B (Methodological)},
  volume={58},
  number  = {1},
  pages={267--288},
  year={1996},
  doi={10.1111/j.2517-6161.1996.tb02080.x}
  }

@article{Weron2014,
  title={{Electricity Price Forecasting: A Review of the State-of-the-Art
          with a Look into the Future}},
  author={Weron, R.},
  journal={International Journal of Forecasting},
  volume={30},
  number={4},
  pages={1030--1081},
  year={2014},
  doi={10.1016/j.ijforecast.2014.08.008}
}

@article{diebold1995paring,
  title={Com paring predictive accu racy},
  author={Diebold, F.X. and Mariano, R.S.},
  journal={Journal of Business and Economic Statistics},
  volume={13},
  number={3},
  pages={253--263},
  year={1995}
}

@article{harvey1997testing,
  title={Testing the equality of prediction mean squared errors},
  author={Harvey, D. and Leybourne, S. and Newbold, P.},
  journal={International Journal of forecasting},
  volume={13},
  number={2},
  pages={281--291},
  year={1997},
  publisher={Elsevier}
}

\break

\begin{appendices} 

\section{Forecast error criteria by hour} \label{app:error-hour}
\setcounter{table}{0}
\renewcommand\thetable{\Alph{section}.\arabic{table}} 

\begin{table}[H]
\centering
\caption{MAE by distribution and hour} \label{tab:mae_hour}
\hspace{-3cm}
\begin{threeparttable}
\begin{tabular}{rcccccc}
  \hline
Hour & Naive & NO & JSU(2) & JSU(3) & JSU(4) & SHASH(2) \\ 
  \hline
1 & $\cellcolor[rgb]{1,0.5,0.5} \underset{(0.82)}{25.59}$ & $\cellcolor[rgb]{0.514,1,0.5} \underset{(0.38)}{11.66}$ & $\cellcolor[rgb]{0.501,1,0.5} \underset{(0.38)}{11.49}$ & $\cellcolor[rgb]{0.5,1,0.5} \underset{(0.38)}{11.47}$ & $\cellcolor[rgb]{0.504,1,0.5} \underset{(0.38)}{11.53}$ & $\cellcolor[rgb]{0.501,1,0.5} \underset{(0.38)}{11.48}$ \\ 
  2 & $\cellcolor[rgb]{1,0.5,0.5} \underset{(0.81)}{26.14}$ & $\cellcolor[rgb]{0.509,1,0.5} \underset{(0.4)}{12.34}$ & $\cellcolor[rgb]{0.505,1,0.5} \underset{(0.4)}{12.28}$ & $\cellcolor[rgb]{0.509,1,0.5} \underset{(0.4)}{12.33}$ & $\cellcolor[rgb]{0.503,1,0.5} \underset{(0.4)}{12.25}$ & $\cellcolor[rgb]{0.5,1,0.5} \underset{(0.39)}{12.21}$ \\ 
  3 & $\cellcolor[rgb]{0.729,1,0.5} \underset{(0.82)}{25.95}$ & $\cellcolor[rgb]{0.503,1,0.5} \underset{(0.41)}{12.76}$ & $\cellcolor[rgb]{1,0.5,0.5} \underset{(58.23)}{71.04}$ & $\cellcolor[rgb]{0.505,1,0.5} \underset{(0.42)}{12.83}$ & $\cellcolor[rgb]{0.504,1,0.5} \underset{(0.41)}{12.79}$ & $\cellcolor[rgb]{0.5,1,0.5} \underset{(0.41)}{12.56}$ \\ 
  4 & $\cellcolor[rgb]{1,0.5,0.5} \underset{(0.83)}{25.95}$ & $\cellcolor[rgb]{0.515,1,0.5} \underset{(0.4)}{13.49}$ & $\cellcolor[rgb]{0.522,1,0.5} \underset{(0.42)}{13.58}$ & $\cellcolor[rgb]{0.53,1,0.5} \underset{(0.42)}{13.68}$ & $\cellcolor[rgb]{0.523,1,0.5} \underset{(0.41)}{13.6}$ & $\cellcolor[rgb]{0.5,1,0.5} \underset{(0.41)}{13.3}$ \\ 
  5 & $\cellcolor[rgb]{1,0.5,0.5} \underset{(0.84)}{25.85}$ & $\cellcolor[rgb]{0.517,1,0.5} \underset{(0.42)}{13.82}$ & $\cellcolor[rgb]{0.528,1,0.5} \underset{(0.43)}{13.94}$ & $\cellcolor[rgb]{0.522,1,0.5} \underset{(0.43)}{13.88}$ & $\cellcolor[rgb]{0.518,1,0.5} \underset{(0.42)}{13.83}$ & $\cellcolor[rgb]{0.5,1,0.5} \underset{(0.42)}{13.61}$ \\ 
  6 & $\cellcolor[rgb]{1,0.5,0.5} \underset{(0.83)}{26.16}$ & $\cellcolor[rgb]{0.522,1,0.5} \underset{(0.44)}{14.7}$ & $\cellcolor[rgb]{0.517,1,0.5} \underset{(0.45)}{14.64}$ & $\cellcolor[rgb]{0.519,1,0.5} \underset{(0.45)}{14.66}$ & $\cellcolor[rgb]{0.514,1,0.5} \underset{(0.44)}{14.61}$ & $\cellcolor[rgb]{0.5,1,0.5} \underset{(0.43)}{14.44}$ \\ 
  7 & $\cellcolor[rgb]{1,0.5,0.5} \underset{(0.82)}{25.88}$ & $\cellcolor[rgb]{0.5,1,0.5} \underset{(0.48)}{17.45}$ & $\cellcolor[rgb]{0.536,1,0.5} \underset{(0.5)}{17.75}$ & $\cellcolor[rgb]{0.532,1,0.5} \underset{(0.49)}{17.71}$ & $\cellcolor[rgb]{0.513,1,0.5} \underset{(0.49)}{17.55}$ & $\cellcolor[rgb]{0.64,1,0.5} \underset{(0.52)}{18.63}$ \\ 
  8 & $\cellcolor[rgb]{1,0.734,0.5} \underset{(0.82)}{24.42}$ & $\cellcolor[rgb]{0.5,1,0.5} \underset{(0.53)}{18.86}$ & $\cellcolor[rgb]{0.598,1,0.5} \underset{(0.55)}{19.57}$ & $\cellcolor[rgb]{1,0.5,0.5} \underset{(6.57)}{26.12}$ & $\cellcolor[rgb]{0.618,1,0.5} \underset{(0.55)}{19.72}$ & $\cellcolor[rgb]{0.583,1,0.5} \underset{(0.55)}{19.46}$ \\ 
  9 & $\cellcolor[rgb]{1,0.5,0.5} \underset{(0.84)}{25.14}$ & $\cellcolor[rgb]{0.5,1,0.5} \underset{(0.59)}{20.66}$ & $\cellcolor[rgb]{0.593,1,0.5} \underset{(0.61)}{21.08}$ & $\cellcolor[rgb]{0.627,1,0.5} \underset{(0.61)}{21.23}$ & $\cellcolor[rgb]{0.609,1,0.5} \underset{(0.61)}{21.15}$ & $\cellcolor[rgb]{0.612,1,0.5} \underset{(0.61)}{21.16}$ \\ 
  10 & $\cellcolor[rgb]{1,0.5,0.5} \underset{(0.83)}{24.86}$ & $\cellcolor[rgb]{0.5,1,0.5} \underset{(0.61)}{22.4}$ & $\cellcolor[rgb]{0.662,1,0.5} \underset{(0.62)}{22.8}$ & $\cellcolor[rgb]{0.71,1,0.5} \underset{(0.62)}{22.92}$ & $\cellcolor[rgb]{0.728,1,0.5} \underset{(0.61)}{22.96}$ & $\cellcolor[rgb]{0.627,1,0.5} \underset{(0.6)}{22.71}$ \\ 
  11 & $\cellcolor[rgb]{1,0.5,0.5} \underset{(0.83)}{25.69}$ & $\cellcolor[rgb]{0.5,1,0.5} \underset{(0.62)}{23.45}$ & $\cellcolor[rgb]{0.864,1,0.5} \underset{(0.65)}{24.27}$ & $\cellcolor[rgb]{0.876,1,0.5} \underset{(0.65)}{24.29}$ & $\cellcolor[rgb]{0.726,1,0.5} \underset{(0.64)}{23.96}$ & $\cellcolor[rgb]{1,0.918,0.5} \underset{(0.71)}{24.75}$ \\ 
  12 & $\cellcolor[rgb]{1,0.5,0.5} \underset{(0.86)}{25.77}$ & $\cellcolor[rgb]{0.584,1,0.5} \underset{(0.63)}{23.51}$ & $\cellcolor[rgb]{0.697,1,0.5} \underset{(0.64)}{23.78}$ & $\cellcolor[rgb]{0.766,1,0.5} \underset{(0.64)}{23.96}$ & $\cellcolor[rgb]{0.57,1,0.5} \underset{(0.63)}{23.47}$ & $\cellcolor[rgb]{0.5,1,0.5} \underset{(0.63)}{23.3}$ \\ 
  13 & $\cellcolor[rgb]{1,0.5,0.5} \underset{(0.88)}{25.59}$ & $\cellcolor[rgb]{0.647,1,0.5} \underset{(0.63)}{22.35}$ & $\cellcolor[rgb]{0.648,1,0.5} \underset{(0.64)}{22.35}$ & $\cellcolor[rgb]{0.639,1,0.5} \underset{(0.65)}{22.32}$ & $\cellcolor[rgb]{0.606,1,0.5} \underset{(0.64)}{22.19}$ & $\cellcolor[rgb]{0.5,1,0.5} \underset{(0.64)}{21.79}$ \\ 
  14 & $\cellcolor[rgb]{1,0.5,0.5} \underset{(0.89)}{25.03}$ & $\cellcolor[rgb]{0.746,1,0.5} \underset{(0.6)}{21.07}$ & $\cellcolor[rgb]{0.702,1,0.5} \underset{(0.61)}{20.84}$ & $\cellcolor[rgb]{0.705,1,0.5} \underset{(0.63)}{20.86}$ & $\cellcolor[rgb]{0.7,1,0.5} \underset{(0.61)}{20.83}$ & $\cellcolor[rgb]{0.5,1,0.5} \underset{(0.59)}{19.78}$ \\ 
  15 & $\cellcolor[rgb]{1,0.5,0.5} \underset{(0.92)}{25.26}$ & $\cellcolor[rgb]{0.693,1,0.5} \underset{(0.61)}{21.03}$ & $\cellcolor[rgb]{0.677,1,0.5} \underset{(0.61)}{20.95}$ & $\cellcolor[rgb]{0.581,1,0.5} \underset{(0.61)}{20.45}$ & $\cellcolor[rgb]{0.588,1,0.5} \underset{(0.6)}{20.48}$ & $\cellcolor[rgb]{0.5,1,0.5} \underset{(0.6)}{20.02}$ \\ 
  16 & $\cellcolor[rgb]{1,0.5,0.5} \underset{(0.92)}{25.27}$ & $\cellcolor[rgb]{0.5,1,0.5} \underset{(0.59)}{20.59}$ & $\cellcolor[rgb]{0.718,1,0.5} \underset{(0.66)}{21.61}$ & $\cellcolor[rgb]{0.634,1,0.5} \underset{(0.64)}{21.22}$ & $\cellcolor[rgb]{0.565,1,0.5} \underset{(0.62)}{20.89}$ & $\cellcolor[rgb]{0.679,1,0.5} \underset{(0.61)}{21.43}$ \\ 
  17 & $\cellcolor[rgb]{1,0.5,0.5} \underset{(0.91)}{25.27}$ & $\cellcolor[rgb]{0.5,1,0.5} \underset{(0.61)}{22.02}$ & $\cellcolor[rgb]{0.738,1,0.5} \underset{(0.67)}{22.79}$ & $\cellcolor[rgb]{0.638,1,0.5} \underset{(0.67)}{22.47}$ & $\cellcolor[rgb]{0.524,1,0.5} \underset{(0.62)}{22.1}$ & $\cellcolor[rgb]{0.58,1,0.5} \underset{(0.63)}{22.28}$ \\ 
  18 & $\cellcolor[rgb]{1,0.5,0.5} \underset{(0.93)}{26.28}$ & $\cellcolor[rgb]{0.683,1,0.5} \underset{(0.66)}{24.42}$ & $\cellcolor[rgb]{0.788,1,0.5} \underset{(0.68)}{24.66}$ & $\cellcolor[rgb]{0.771,1,0.5} \underset{(0.68)}{24.62}$ & $\cellcolor[rgb]{0.5,1,0.5} \underset{(0.65)}{24}$ & $\cellcolor[rgb]{0.607,1,0.5} \underset{(0.66)}{24.24}$ \\ 
  19 & $\cellcolor[rgb]{0.51,1,0.5} \underset{(0.88)}{26.55}$ & $\cellcolor[rgb]{0.5,1,0.5} \underset{(0.69)}{26.54}$ & $\cellcolor[rgb]{1,0.5,0.5} \underset{(0.69)}{27.89}$ & $\cellcolor[rgb]{1,0.566,0.5} \underset{(0.69)}{27.8}$ & $\cellcolor[rgb]{1,0.615,0.5} \underset{(0.68)}{27.73}$ & $\cellcolor[rgb]{1,0.58,0.5} \underset{(0.68)}{27.78}$ \\ 
  20 & $\cellcolor[rgb]{1,0.647,0.5} \underset{(0.87)}{26.44}$ & $\cellcolor[rgb]{1,0.5,0.5} \underset{(0.68)}{26.75}$ & $\cellcolor[rgb]{0.555,1,0.5} \underset{(0.67)}{24.75}$ & $\cellcolor[rgb]{0.5,1,0.5} \underset{(0.67)}{24.63}$ & $\cellcolor[rgb]{0.713,1,0.5} \underset{(0.7)}{25.08}$ & $\cellcolor[rgb]{0.858,1,0.5} \underset{(0.68)}{25.39}$ \\ 
  21 & $\cellcolor[rgb]{1,0.5,0.5} \underset{(0.83)}{25.46}$ & $\cellcolor[rgb]{0.693,1,0.5} \underset{(0.61)}{20.85}$ & $\cellcolor[rgb]{0.506,1,0.5} \underset{(0.6)}{19.78}$ & $\cellcolor[rgb]{0.5,1,0.5} \underset{(0.6)}{19.75}$ & $\cellcolor[rgb]{0.601,1,0.5} \underset{(0.63)}{20.32}$ & $\cellcolor[rgb]{0.51,1,0.5} \underset{(0.6)}{19.8}$ \\ 
  22 & $\cellcolor[rgb]{1,0.5,0.5} \underset{(0.78)}{23.62}$ & $\cellcolor[rgb]{0.556,1,0.5} \underset{(0.56)}{18.62}$ & $\cellcolor[rgb]{0.5,1,0.5} \underset{(0.56)}{18.32}$ & $\cellcolor[rgb]{0.549,1,0.5} \underset{(0.58)}{18.58}$ & $\cellcolor[rgb]{0.554,1,0.5} \underset{(0.58)}{18.61}$ & $\cellcolor[rgb]{0.522,1,0.5} \underset{(0.57)}{18.44}$ \\ 
  23 & $\cellcolor[rgb]{1,0.5,0.5} \underset{(0.77)}{21.91}$ & $\cellcolor[rgb]{0.675,1,0.5} \underset{(0.54)}{17.86}$ & $\cellcolor[rgb]{0.514,1,0.5} \underset{(0.55)}{17.07}$ & $\cellcolor[rgb]{0.5,1,0.5} \underset{(0.55)}{17.01}$ & $\cellcolor[rgb]{0.55,1,0.5} \underset{(0.56)}{17.25}$ & $\cellcolor[rgb]{0.601,1,0.5} \underset{(0.56)}{17.5}$ \\ 
  24 & $\cellcolor[rgb]{1,0.5,0.5} \underset{(0.78)}{22.82}$ & $\cellcolor[rgb]{0.775,1,0.5} \underset{(0.53)}{19.37}$ & $\cellcolor[rgb]{0.511,1,0.5} \underset{(0.54)}{18.11}$ & $\cellcolor[rgb]{0.516,1,0.5} \underset{(0.54)}{18.13}$ & $\cellcolor[rgb]{0.562,1,0.5} \underset{(0.55)}{18.35}$ & $\cellcolor[rgb]{0.5,1,0.5} \underset{(0.54)}{18.06}$ \\ 
   \hline
\end{tabular}
\begin{tablenotes}
\footnotesize
\item Note: NO, JSU, and SHASH denote the Normal, Johnson's SU, and Sinh–Arcsinh distributions, respectively. The number in parentheses next to each distribution name indicates how many distribution moments are modelled as non-constant, i.e., as functions of the explanatory variables. For example, JSU(2) means that the first two moments are non-constant. Parentheses beneath the error measures contain the corresponding standard deviations. A heat map is used to indicate lower (green) and higher (red) forecast errors.
\end{tablenotes}
\end{threeparttable}
\end{table}

\newpage

\begin{table}[H]
\centering
\caption{PB $\tau=0.99$ by distribution and hour}\label{tab:PB99_hour}
\hspace{-3cm}
\begin{threeparttable}
\begin{tabular}{rcccccc}
  \hline
Hour & Naive & NO & JSU(2) & JSU(3) & JSU(4) & SHASH(2) \\ 
  \hline
1 & $\cellcolor[rgb]{1,0.5,0.5} \underset{(0.22)}{1.83}$ & $\cellcolor[rgb]{0.702,1,0.5} \underset{(0.11)}{0.96}$ & $\cellcolor[rgb]{0.598,1,0.5} \underset{(0.09)}{0.85}$ & $\cellcolor[rgb]{0.592,1,0.5} \underset{(0.08)}{0.85}$ & $\cellcolor[rgb]{0.5,1,0.5} \underset{(0.04)}{0.75}$ & $\cellcolor[rgb]{0.526,1,0.5} \underset{(0.08)}{0.78}$ \\ 
  2 & $\cellcolor[rgb]{1,0.5,0.5} \underset{(0.21)}{1.75}$ & $\cellcolor[rgb]{0.57,1,0.5} \underset{(0.08)}{0.79}$ & $\cellcolor[rgb]{0.533,1,0.5} \underset{(0.08)}{0.75}$ & $\cellcolor[rgb]{0.506,1,0.5} \underset{(0.07)}{0.72}$ & $\cellcolor[rgb]{0.5,1,0.5} \underset{(0.05)}{0.71}$ & $\cellcolor[rgb]{0.563,1,0.5} \underset{(0.07)}{0.78}$ \\ 
  3 & $\cellcolor[rgb]{1,0.5,0.5} \underset{(0.2)}{1.69}$ & $\cellcolor[rgb]{0.537,1,0.5} \underset{(0.08)}{0.78}$ & $\cellcolor[rgb]{1,0.898,0.5} \underset{(0.6)}{1.32}$ & $\cellcolor[rgb]{0.517,1,0.5} \underset{(0.08)}{0.76}$ & $\cellcolor[rgb]{0.5,1,0.5} \underset{(0.05)}{0.75}$ & $\cellcolor[rgb]{0.564,1,0.5} \underset{(0.07)}{0.81}$ \\ 
  4 & $\cellcolor[rgb]{1,0.5,0.5} \underset{(0.19)}{1.68}$ & $\cellcolor[rgb]{0.543,1,0.5} \underset{(0.07)}{0.72}$ & $\cellcolor[rgb]{0.5,1,0.5} \underset{(0.05)}{0.67}$ & $\cellcolor[rgb]{0.524,1,0.5} \underset{(0.06)}{0.7}$ & $\cellcolor[rgb]{0.501,1,0.5} \underset{(0.05)}{0.67}$ & $\cellcolor[rgb]{0.544,1,0.5} \underset{(0.06)}{0.72}$ \\ 
  5 & $\cellcolor[rgb]{1,0.5,0.5} \underset{(0.18)}{1.72}$ & $\cellcolor[rgb]{0.503,1,0.5} \underset{(0.07)}{0.7}$ & $\cellcolor[rgb]{0.536,1,0.5} \underset{(0.06)}{0.73}$ & $\cellcolor[rgb]{0.514,1,0.5} \underset{(0.06)}{0.71}$ & $\cellcolor[rgb]{0.5,1,0.5} \underset{(0.05)}{0.69}$ & $\cellcolor[rgb]{0.605,1,0.5} \underset{(0.05)}{0.8}$ \\ 
  6 & $\cellcolor[rgb]{1,0.5,0.5} \underset{(0.18)}{1.69}$ & $\cellcolor[rgb]{0.5,1,0.5} \underset{(0.08)}{0.76}$ & $\cellcolor[rgb]{0.545,1,0.5} \underset{(0.07)}{0.81}$ & $\cellcolor[rgb]{0.542,1,0.5} \underset{(0.08)}{0.8}$ & $\cellcolor[rgb]{0.531,1,0.5} \underset{(0.05)}{0.79}$ & $\cellcolor[rgb]{0.618,1,0.5} \underset{(0.05)}{0.87}$ \\ 
  7 & $\cellcolor[rgb]{1,0.5,0.5} \underset{(0.19)}{1.69}$ & $\cellcolor[rgb]{0.5,1,0.5} \underset{(0.09)}{0.72}$ & $\cellcolor[rgb]{0.593,1,0.5} \underset{(0.07)}{0.81}$ & $\cellcolor[rgb]{0.587,1,0.5} \underset{(0.07)}{0.8}$ & $\cellcolor[rgb]{0.76,1,0.5} \underset{(0.06)}{0.97}$ & $\cellcolor[rgb]{0.605,1,0.5} \underset{(0.08)}{0.82}$ \\ 
  8 & $\cellcolor[rgb]{0.556,1,0.5} \underset{(0.17)}{1.61}$ & $\cellcolor[rgb]{0.505,1,0.5} \underset{(0.15)}{1.27}$ & $\cellcolor[rgb]{0.5,1,0.5} \underset{(0.13)}{1.25}$ & $\cellcolor[rgb]{1,0.5,0.5} \underset{(6.46)}{7.75}$ & $\cellcolor[rgb]{0.526,1,0.5} \underset{(0.13)}{1.42}$ & $\cellcolor[rgb]{0.5,1,0.5} \underset{(0.12)}{1.24}$ \\ 
  9 & $\cellcolor[rgb]{1,0.5,0.5} \underset{(0.2)}{1.73}$ & $\cellcolor[rgb]{0.549,1,0.5} \underset{(0.16)}{1.29}$ & $\cellcolor[rgb]{0.5,1,0.5} \underset{(0.13)}{1.26}$ & $\cellcolor[rgb]{0.636,1,0.5} \underset{(0.13)}{1.33}$ & $\cellcolor[rgb]{0.928,1,0.5} \underset{(0.17)}{1.46}$ & $\cellcolor[rgb]{0.852,1,0.5} \underset{(0.14)}{1.43}$ \\ 
  10 & $\cellcolor[rgb]{1,0.5,0.5} \underset{(0.2)}{1.6}$ & $\cellcolor[rgb]{0.844,1,0.5} \underset{(0.16)}{1.32}$ & $\cellcolor[rgb]{0.5,1,0.5} \underset{(0.11)}{1.17}$ & $\cellcolor[rgb]{0.85,1,0.5} \underset{(0.15)}{1.32}$ & $\cellcolor[rgb]{0.892,1,0.5} \underset{(0.12)}{1.34}$ & $\cellcolor[rgb]{0.726,1,0.5} \underset{(0.11)}{1.27}$ \\ 
  11 & $\cellcolor[rgb]{1,0.5,0.5} \underset{(0.21)}{1.72}$ & $\cellcolor[rgb]{0.937,1,0.5} \underset{(0.14)}{1.3}$ & $\cellcolor[rgb]{0.5,1,0.5} \underset{(0.1)}{0.98}$ & $\cellcolor[rgb]{0.576,1,0.5} \underset{(0.11)}{1.04}$ & $\cellcolor[rgb]{0.543,1,0.5} \underset{(0.07)}{1.02}$ & $\cellcolor[rgb]{1,0.893,0.5} \underset{(0.29)}{1.43}$ \\ 
  12 & $\cellcolor[rgb]{1,0.5,0.5} \underset{(0.21)}{1.7}$ & $\cellcolor[rgb]{0.52,1,0.5} \underset{(0.1)}{0.99}$ & $\cellcolor[rgb]{0.5,1,0.5} \underset{(0.08)}{0.98}$ & $\cellcolor[rgb]{0.5,1,0.5} \underset{(0.08)}{0.97}$ & $\cellcolor[rgb]{0.512,1,0.5} \underset{(0.07)}{0.98}$ & $\cellcolor[rgb]{0.529,1,0.5} \underset{(0.07)}{1}$ \\ 
  13 & $\cellcolor[rgb]{1,0.5,0.5} \underset{(0.21)}{1.72}$ & $\cellcolor[rgb]{0.5,1,0.5} \underset{(0.1)}{1}$ & $\cellcolor[rgb]{0.884,1,0.5} \underset{(0.09)}{1.27}$ & $\cellcolor[rgb]{1,0.923,0.5} \underset{(0.11)}{1.41}$ & $\cellcolor[rgb]{0.588,1,0.5} \underset{(0.09)}{1.06}$ & $\cellcolor[rgb]{1,0.838,0.5} \underset{(0.1)}{1.47}$ \\ 
  14 & $\cellcolor[rgb]{1,0.5,0.5} \underset{(0.23)}{1.82}$ & $\cellcolor[rgb]{0.5,1,0.5} \underset{(0.11)}{1.03}$ & $\cellcolor[rgb]{0.792,1,0.5} \underset{(0.1)}{1.26}$ & $\cellcolor[rgb]{0.933,1,0.5} \underset{(0.11)}{1.37}$ & $\cellcolor[rgb]{0.504,1,0.5} \underset{(0.09)}{1.03}$ & $\cellcolor[rgb]{1,0.885,0.5} \underset{(0.13)}{1.52}$ \\ 
  15 & $\cellcolor[rgb]{1,0.5,0.5} \underset{(0.23)}{1.9}$ & $\cellcolor[rgb]{0.5,1,0.5} \underset{(0.1)}{0.94}$ & $\cellcolor[rgb]{1,0.762,0.5} \underset{(0.14)}{1.65}$ & $\cellcolor[rgb]{1,0.684,0.5} \underset{(0.14)}{1.72}$ & $\cellcolor[rgb]{0.904,1,0.5} \underset{(0.12)}{1.33}$ & $\cellcolor[rgb]{1,0.83,0.5} \underset{(0.13)}{1.58}$ \\ 
  16 & $\cellcolor[rgb]{1,0.721,0.5} \underset{(0.23)}{1.94}$ & $\cellcolor[rgb]{0.5,1,0.5} \underset{(0.12)}{1.16}$ & $\cellcolor[rgb]{1,0.729,0.5} \underset{(0.22)}{1.93}$ & $\cellcolor[rgb]{1,0.928,0.5} \underset{(0.15)}{1.73}$ & $\cellcolor[rgb]{0.895,1,0.5} \underset{(0.18)}{1.56}$ & $\cellcolor[rgb]{1,0.5,0.5} \underset{(0.18)}{2.16}$ \\ 
  17 & $\cellcolor[rgb]{1,0.64,0.5} \underset{(0.21)}{1.9}$ & $\cellcolor[rgb]{0.5,1,0.5} \underset{(0.13)}{1.19}$ & $\cellcolor[rgb]{1,0.782,0.5} \underset{(0.15)}{1.78}$ & $\cellcolor[rgb]{1,0.877,0.5} \underset{(0.14)}{1.7}$ & $\cellcolor[rgb]{0.72,1,0.5} \underset{(0.12)}{1.37}$ & $\cellcolor[rgb]{1,0.5,0.5} \underset{(0.18)}{2.01}$ \\ 
  18 & $\cellcolor[rgb]{1,0.5,0.5} \underset{(0.24)}{2.01}$ & $\cellcolor[rgb]{0.933,1,0.5} \underset{(0.19)}{1.71}$ & $\cellcolor[rgb]{0.699,1,0.5} \underset{(0.16)}{1.58}$ & $\cellcolor[rgb]{0.5,1,0.5} \underset{(0.16)}{1.48}$ & $\cellcolor[rgb]{0.754,1,0.5} \underset{(0.2)}{1.61}$ & $\cellcolor[rgb]{0.519,1,0.5} \underset{(0.14)}{1.49}$ \\ 
  19 & $\cellcolor[rgb]{0.5,1,0.5} \underset{(0.25)}{1.93}$ & $\cellcolor[rgb]{1,0.5,0.5} \underset{(0.3)}{3.29}$ & $\cellcolor[rgb]{1,0.684,0.5} \underset{(0.25)}{3.04}$ & $\cellcolor[rgb]{1,0.909,0.5} \underset{(0.23)}{2.73}$ & $\cellcolor[rgb]{1,0.778,0.5} \underset{(0.25)}{2.91}$ & $\cellcolor[rgb]{0.999,1,0.5} \underset{(0.23)}{2.6}$ \\ 
  20 & $\cellcolor[rgb]{0.5,1,0.5} \underset{(0.26)}{1.93}$ & $\cellcolor[rgb]{1,0.5,0.5} \underset{(0.42)}{5.79}$ & $\cellcolor[rgb]{0.633,1,0.5} \underset{(0.27)}{2.44}$ & $\cellcolor[rgb]{0.646,1,0.5} \underset{(0.28)}{2.49}$ & $\cellcolor[rgb]{0.625,1,0.5} \underset{(0.32)}{2.41}$ & $\cellcolor[rgb]{0.63,1,0.5} \underset{(0.25)}{2.43}$ \\ 
  21 & $\cellcolor[rgb]{0.54,1,0.5} \underset{(0.23)}{1.85}$ & $\cellcolor[rgb]{1,0.5,0.5} \underset{(0.35)}{3.1}$ & $\cellcolor[rgb]{0.805,1,0.5} \underset{(0.25)}{2.19}$ & $\cellcolor[rgb]{0.807,1,0.5} \underset{(0.25)}{2.2}$ & $\cellcolor[rgb]{0.5,1,0.5} \underset{(0.19)}{1.8}$ & $\cellcolor[rgb]{0.818,1,0.5} \underset{(0.23)}{2.21}$ \\ 
  22 & $\cellcolor[rgb]{0.66,1,0.5} \underset{(0.22)}{1.67}$ & $\cellcolor[rgb]{1,0.5,0.5} \underset{(0.32)}{2.85}$ & $\cellcolor[rgb]{0.686,1,0.5} \underset{(0.18)}{1.71}$ & $\cellcolor[rgb]{0.774,1,0.5} \underset{(0.21)}{1.83}$ & $\cellcolor[rgb]{0.5,1,0.5} \underset{(0.14)}{1.45}$ & $\cellcolor[rgb]{0.904,1,0.5} \underset{(0.18)}{2.02}$ \\ 
  23 & $\cellcolor[rgb]{0.751,1,0.5} \underset{(0.22)}{1.58}$ & $\cellcolor[rgb]{1,0.5,0.5} \underset{(0.29)}{2.09}$ & $\cellcolor[rgb]{0.603,1,0.5} \underset{(0.19)}{1.48}$ & $\cellcolor[rgb]{0.781,1,0.5} \underset{(0.21)}{1.6}$ & $\cellcolor[rgb]{0.5,1,0.5} \underset{(0.18)}{1.41}$ & $\cellcolor[rgb]{0.626,1,0.5} \underset{(0.16)}{1.49}$ \\ 
  24 & $\cellcolor[rgb]{1,0.5,0.5} \underset{(0.23)}{1.68}$ & $\cellcolor[rgb]{1,0.938,0.5} \underset{(0.19)}{1.43}$ & $\cellcolor[rgb]{0.564,1,0.5} \underset{(0.11)}{1.15}$ & $\cellcolor[rgb]{0.605,1,0.5} \underset{(0.12)}{1.17}$ & $\cellcolor[rgb]{1,0.905,0.5} \underset{(0.2)}{1.45}$ & $\cellcolor[rgb]{0.5,1,0.5} \underset{(0.09)}{1.11}$ \\ 
   \hline
\end{tabular}
\begin{tablenotes}
\footnotesize
\item Note: NO, JSU, and SHASH denote the Normal, Johnson's SU, and Sinh–Arcsinh distributions, respectively. The number in parentheses next to each distribution name indicates how many distribution moments are modelled as non-constant, i.e., as functions of the explanatory variables. For example, JSU(2) means that the first two moments are non-constant. Parentheses beneath the error measures contain the corresponding standard deviations. A heat map is used to indicate lower (green) and higher (red) forecast errors.
\end{tablenotes}
\end{threeparttable}
\end{table}

\newpage

\begin{table}[H]
\centering
\caption{PB $\tau=0.95$ by distribution and hour}\label{tab:PB95_hour}
\hspace{-3cm}
\begin{threeparttable}
\begin{tabular}{rcccccc}
  \hline
Hour & Naive & NO & JSU(2) & JSU(3) & JSU(4) & SHASH(2) \\ 
  \hline
1 & $\cellcolor[rgb]{1,0.5,0.5} \underset{(0.29)}{4.83}$ & $\cellcolor[rgb]{0.565,1,0.5} \underset{(0.14)}{2.18}$ & $\cellcolor[rgb]{0.586,1,0.5} \underset{(0.14)}{2.24}$ & $\cellcolor[rgb]{0.575,1,0.5} \underset{(0.13)}{2.21}$ & $\cellcolor[rgb]{0.5,1,0.5} \underset{(0.09)}{2}$ & $\cellcolor[rgb]{0.573,1,0.5} \underset{(0.13)}{2.21}$ \\ 
  2 & $\cellcolor[rgb]{1,0.5,0.5} \underset{(0.28)}{4.83}$ & $\cellcolor[rgb]{0.534,1,0.5} \underset{(0.12)}{2.05}$ & $\cellcolor[rgb]{0.538,1,0.5} \underset{(0.12)}{2.06}$ & $\cellcolor[rgb]{0.532,1,0.5} \underset{(0.12)}{2.04}$ & $\cellcolor[rgb]{0.5,1,0.5} \underset{(0.1)}{1.95}$ & $\cellcolor[rgb]{0.614,1,0.5} \underset{(0.12)}{2.28}$ \\ 
  3 & $\cellcolor[rgb]{1,0.593,0.5} \underset{(0.27)}{4.77}$ & $\cellcolor[rgb]{0.537,1,0.5} \underset{(0.11)}{2.18}$ & $\cellcolor[rgb]{1,0.5,0.5} \underset{(2.91)}{5.04}$ & $\cellcolor[rgb]{0.519,1,0.5} \underset{(0.11)}{2.12}$ & $\cellcolor[rgb]{0.5,1,0.5} \underset{(0.09)}{2.07}$ & $\cellcolor[rgb]{0.617,1,0.5} \underset{(0.11)}{2.41}$ \\ 
  4 & $\cellcolor[rgb]{1,0.5,0.5} \underset{(0.27)}{4.8}$ & $\cellcolor[rgb]{0.523,1,0.5} \underset{(0.1)}{2.09}$ & $\cellcolor[rgb]{0.536,1,0.5} \underset{(0.09)}{2.13}$ & $\cellcolor[rgb]{0.553,1,0.5} \underset{(0.1)}{2.18}$ & $\cellcolor[rgb]{0.5,1,0.5} \underset{(0.09)}{2.03}$ & $\cellcolor[rgb]{0.582,1,0.5} \underset{(0.09)}{2.25}$ \\ 
  5 & $\cellcolor[rgb]{1,0.5,0.5} \underset{(0.26)}{4.87}$ & $\cellcolor[rgb]{0.517,1,0.5} \underset{(0.1)}{2.14}$ & $\cellcolor[rgb]{0.557,1,0.5} \underset{(0.1)}{2.25}$ & $\cellcolor[rgb]{0.527,1,0.5} \underset{(0.1)}{2.17}$ & $\cellcolor[rgb]{0.5,1,0.5} \underset{(0.1)}{2.1}$ & $\cellcolor[rgb]{0.663,1,0.5} \underset{(0.1)}{2.55}$ \\ 
  6 & $\cellcolor[rgb]{1,0.5,0.5} \underset{(0.26)}{4.85}$ & $\cellcolor[rgb]{0.5,1,0.5} \underset{(0.11)}{2.28}$ & $\cellcolor[rgb]{0.537,1,0.5} \underset{(0.12)}{2.37}$ & $\cellcolor[rgb]{0.525,1,0.5} \underset{(0.12)}{2.34}$ & $\cellcolor[rgb]{0.545,1,0.5} \underset{(0.12)}{2.39}$ & $\cellcolor[rgb]{0.699,1,0.5} \underset{(0.11)}{2.79}$ \\ 
  7 & $\cellcolor[rgb]{1,0.5,0.5} \underset{(0.26)}{4.78}$ & $\cellcolor[rgb]{0.5,1,0.5} \underset{(0.13)}{2.36}$ & $\cellcolor[rgb]{0.589,1,0.5} \underset{(0.14)}{2.58}$ & $\cellcolor[rgb]{0.582,1,0.5} \underset{(0.13)}{2.56}$ & $\cellcolor[rgb]{0.619,1,0.5} \underset{(0.12)}{2.65}$ & $\cellcolor[rgb]{0.615,1,0.5} \underset{(0.13)}{2.64}$ \\ 
  8 & $\cellcolor[rgb]{0.738,1,0.5} \underset{(0.25)}{4.62}$ & $\cellcolor[rgb]{0.5,1,0.5} \underset{(0.2)}{3.03}$ & $\cellcolor[rgb]{0.56,1,0.5} \underset{(0.2)}{3.44}$ & $\cellcolor[rgb]{1,0.5,0.5} \underset{(6.21)}{9.71}$ & $\cellcolor[rgb]{0.585,1,0.5} \underset{(0.21)}{3.6}$ & $\cellcolor[rgb]{0.555,1,0.5} \underset{(0.19)}{3.4}$ \\ 
  9 & $\cellcolor[rgb]{1,0.5,0.5} \underset{(0.27)}{4.8}$ & $\cellcolor[rgb]{0.5,1,0.5} \underset{(0.21)}{3.24}$ & $\cellcolor[rgb]{0.747,1,0.5} \underset{(0.2)}{3.63}$ & $\cellcolor[rgb]{0.782,1,0.5} \underset{(0.2)}{3.68}$ & $\cellcolor[rgb]{0.776,1,0.5} \underset{(0.22)}{3.67}$ & $\cellcolor[rgb]{0.981,1,0.5} \underset{(0.21)}{3.99}$ \\ 
  10 & $\cellcolor[rgb]{1,0.5,0.5} \underset{(0.26)}{4.57}$ & $\cellcolor[rgb]{0.5,1,0.5} \underset{(0.19)}{3.21}$ & $\cellcolor[rgb]{0.531,1,0.5} \underset{(0.16)}{3.25}$ & $\cellcolor[rgb]{0.65,1,0.5} \underset{(0.19)}{3.41}$ & $\cellcolor[rgb]{0.582,1,0.5} \underset{(0.18)}{3.32}$ & $\cellcolor[rgb]{0.719,1,0.5} \underset{(0.15)}{3.51}$ \\ 
  11 & $\cellcolor[rgb]{1,0.5,0.5} \underset{(0.27)}{4.81}$ & $\cellcolor[rgb]{0.662,1,0.5} \underset{(0.18)}{3.27}$ & $\cellcolor[rgb]{0.548,1,0.5} \underset{(0.16)}{3.06}$ & $\cellcolor[rgb]{0.569,1,0.5} \underset{(0.17)}{3.1}$ & $\cellcolor[rgb]{0.5,1,0.5} \underset{(0.15)}{2.97}$ & $\cellcolor[rgb]{0.837,1,0.5} \underset{(0.31)}{3.59}$ \\ 
  12 & $\cellcolor[rgb]{1,0.5,0.5} \underset{(0.28)}{4.89}$ & $\cellcolor[rgb]{0.5,1,0.5} \underset{(0.15)}{2.92}$ & $\cellcolor[rgb]{0.625,1,0.5} \underset{(0.13)}{3.17}$ & $\cellcolor[rgb]{0.641,1,0.5} \underset{(0.14)}{3.2}$ & $\cellcolor[rgb]{0.545,1,0.5} \underset{(0.14)}{3.01}$ & $\cellcolor[rgb]{0.688,1,0.5} \underset{(0.14)}{3.29}$ \\ 
  13 & $\cellcolor[rgb]{1,0.5,0.5} \underset{(0.28)}{4.91}$ & $\cellcolor[rgb]{0.5,1,0.5} \underset{(0.14)}{2.92}$ & $\cellcolor[rgb]{1,0.917,0.5} \underset{(0.2)}{4.08}$ & $\cellcolor[rgb]{1,0.718,0.5} \underset{(0.26)}{4.48}$ & $\cellcolor[rgb]{0.614,1,0.5} \underset{(0.15)}{3.14}$ & $\cellcolor[rgb]{1,0.628,0.5} \underset{(0.27)}{4.66}$ \\ 
  14 & $\cellcolor[rgb]{1,0.5,0.5} \underset{(0.29)}{4.98}$ & $\cellcolor[rgb]{0.5,1,0.5} \underset{(0.15)}{2.93}$ & $\cellcolor[rgb]{1,0.99,0.5} \underset{(0.21)}{3.97}$ & $\cellcolor[rgb]{1,0.892,0.5} \underset{(0.23)}{4.17}$ & $\cellcolor[rgb]{0.559,1,0.5} \underset{(0.15)}{3.05}$ & $\cellcolor[rgb]{1,0.706,0.5} \underset{(0.29)}{4.55}$ \\ 
  15 & $\cellcolor[rgb]{1,0.5,0.5} \underset{(0.3)}{5.1}$ & $\cellcolor[rgb]{0.5,1,0.5} \underset{(0.14)}{2.84}$ & $\cellcolor[rgb]{1,0.683,0.5} \underset{(0.29)}{4.69}$ & $\cellcolor[rgb]{1,0.619,0.5} \underset{(0.31)}{4.83}$ & $\cellcolor[rgb]{0.755,1,0.5} \underset{(0.17)}{3.41}$ & $\cellcolor[rgb]{1,0.65,0.5} \underset{(0.29)}{4.76}$ \\ 
  16 & $\cellcolor[rgb]{1,0.8,0.5} \underset{(0.3)}{5.19}$ & $\cellcolor[rgb]{0.5,1,0.5} \underset{(0.16)}{3.07}$ & $\cellcolor[rgb]{1,0.816,0.5} \underset{(0.31)}{5.14}$ & $\cellcolor[rgb]{1,0.927,0.5} \underset{(0.26)}{4.81}$ & $\cellcolor[rgb]{0.748,1,0.5} \underset{(0.23)}{3.82}$ & $\cellcolor[rgb]{1,0.5,0.5} \underset{(0.4)}{6.1}$ \\ 
  17 & $\cellcolor[rgb]{1,0.867,0.5} \underset{(0.29)}{5.16}$ & $\cellcolor[rgb]{0.5,1,0.5} \underset{(0.18)}{3.26}$ & $\cellcolor[rgb]{1,0.797,0.5} \underset{(0.29)}{5.37}$ & $\cellcolor[rgb]{1,0.818,0.5} \underset{(0.29)}{5.3}$ & $\cellcolor[rgb]{0.789,1,0.5} \underset{(0.23)}{4.13}$ & $\cellcolor[rgb]{1,0.5,0.5} \underset{(0.45)}{6.26}$ \\ 
  18 & $\cellcolor[rgb]{1,0.5,0.5} \underset{(0.31)}{5.29}$ & $\cellcolor[rgb]{0.5,1,0.5} \underset{(0.24)}{4.19}$ & $\cellcolor[rgb]{1,0.768,0.5} \underset{(0.25)}{4.99}$ & $\cellcolor[rgb]{1,0.902,0.5} \underset{(0.25)}{4.85}$ & $\cellcolor[rgb]{0.982,1,0.5} \underset{(0.29)}{4.72}$ & $\cellcolor[rgb]{1,0.863,0.5} \underset{(0.26)}{4.89}$ \\ 
  19 & $\cellcolor[rgb]{0.5,1,0.5} \underset{(0.31)}{5.04}$ & $\cellcolor[rgb]{0.992,1,0.5} \underset{(0.35)}{6.25}$ & $\cellcolor[rgb]{1,0.5,0.5} \underset{(0.36)}{7.5}$ & $\cellcolor[rgb]{1,0.62,0.5} \underset{(0.34)}{7.2}$ & $\cellcolor[rgb]{1,0.57,0.5} \underset{(0.36)}{7.33}$ & $\cellcolor[rgb]{1,0.653,0.5} \underset{(0.35)}{7.12}$ \\ 
  20 & $\cellcolor[rgb]{0.5,1,0.5} \underset{(0.31)}{4.96}$ & $\cellcolor[rgb]{1,0.5,0.5} \underset{(0.46)}{8.45}$ & $\cellcolor[rgb]{0.674,1,0.5} \underset{(0.36)}{5.56}$ & $\cellcolor[rgb]{0.635,1,0.5} \underset{(0.37)}{5.43}$ & $\cellcolor[rgb]{0.657,1,0.5} \underset{(0.41)}{5.51}$ & $\cellcolor[rgb]{0.713,1,0.5} \underset{(0.34)}{5.7}$ \\ 
  21 & $\cellcolor[rgb]{0.947,1,0.5} \underset{(0.29)}{4.82}$ & $\cellcolor[rgb]{1,0.5,0.5} \underset{(0.38)}{5.39}$ & $\cellcolor[rgb]{0.904,1,0.5} \underset{(0.32)}{4.77}$ & $\cellcolor[rgb]{0.896,1,0.5} \underset{(0.32)}{4.76}$ & $\cellcolor[rgb]{0.5,1,0.5} \underset{(0.29)}{4.35}$ & $\cellcolor[rgb]{1,0.876,0.5} \underset{(0.3)}{5}$ \\ 
  22 & $\cellcolor[rgb]{1,0.957,0.5} \underset{(0.28)}{4.57}$ & $\cellcolor[rgb]{1,0.937,0.5} \underset{(0.35)}{4.6}$ & $\cellcolor[rgb]{0.788,1,0.5} \underset{(0.27)}{4.23}$ & $\cellcolor[rgb]{0.866,1,0.5} \underset{(0.29)}{4.33}$ & $\cellcolor[rgb]{0.5,1,0.5} \underset{(0.25)}{3.84}$ & $\cellcolor[rgb]{1,0.5,0.5} \underset{(0.29)}{5.19}$ \\ 
  23 & $\cellcolor[rgb]{1,0.5,0.5} \underset{(0.27)}{4.29}$ & $\cellcolor[rgb]{0.739,1,0.5} \underset{(0.31)}{3.71}$ & $\cellcolor[rgb]{0.5,1,0.5} \underset{(0.26)}{3.53}$ & $\cellcolor[rgb]{0.578,1,0.5} \underset{(0.28)}{3.59}$ & $\cellcolor[rgb]{0.563,1,0.5} \underset{(0.27)}{3.58}$ & $\cellcolor[rgb]{0.827,1,0.5} \underset{(0.24)}{3.78}$ \\ 
  24 & $\cellcolor[rgb]{1,0.5,0.5} \underset{(0.28)}{4.41}$ & $\cellcolor[rgb]{0.54,1,0.5} \underset{(0.23)}{3.29}$ & $\cellcolor[rgb]{0.5,1,0.5} \underset{(0.19)}{3.24}$ & $\cellcolor[rgb]{0.536,1,0.5} \underset{(0.19)}{3.28}$ & $\cellcolor[rgb]{0.509,1,0.5} \underset{(0.25)}{3.25}$ & $\cellcolor[rgb]{0.627,1,0.5} \underset{(0.16)}{3.39}$ \\ 
   \hline
\end{tabular}
\begin{tablenotes}
\footnotesize
\item Note: NO, JSU, and SHASH denote the Normal, Johnson's SU, and Sinh–Arcsinh distributions, respectively. The number in parentheses next to each distribution name indicates how many distribution moments are modelled as non-constant, i.e., as functions of the explanatory variables. For example, JSU(2) means that the first two moments are non-constant. Parentheses beneath the error measures contain the corresponding standard deviations. A heat map is used to indicate lower (green) and higher (red) forecast errors.
\end{tablenotes}
\end{threeparttable}
\end{table}

\newpage

\begin{table}[H]
\centering
\caption{PB $\tau=0.05$ by distribution and hour}\label{tab:PB5_hour}
\hspace{-3cm}
\begin{threeparttable}
\begin{tabular}{rcccccc}
  \hline
Hour & Naive & NO & JSU(2) & JSU(3) & JSU(4) & SHASH(2) \\ 
  \hline
1 & $\cellcolor[rgb]{1,0.5,0.5} \underset{(0.26)}{4.62}$ & $\cellcolor[rgb]{0.525,1,0.5} \underset{(0.15)}{2.07}$ & $\cellcolor[rgb]{0.504,1,0.5} \underset{(0.11)}{2.01}$ & $\cellcolor[rgb]{0.5,1,0.5} \underset{(0.11)}{2}$ & $\cellcolor[rgb]{0.512,1,0.5} \underset{(0.1)}{2.03}$ & $\cellcolor[rgb]{0.563,1,0.5} \underset{(0.12)}{2.17}$ \\ 
  2 & $\cellcolor[rgb]{1,0.5,0.5} \underset{(0.23)}{4.49}$ & $\cellcolor[rgb]{0.523,1,0.5} \underset{(0.15)}{2.34}$ & $\cellcolor[rgb]{0.5,1,0.5} \underset{(0.13)}{2.29}$ & $\cellcolor[rgb]{0.55,1,0.5} \underset{(0.14)}{2.4}$ & $\cellcolor[rgb]{0.519,1,0.5} \underset{(0.13)}{2.34}$ & $\cellcolor[rgb]{0.624,1,0.5} \underset{(0.12)}{2.57}$ \\ 
  3 & $\cellcolor[rgb]{0.538,1,0.5} \underset{(0.22)}{4.47}$ & $\cellcolor[rgb]{0.5,1,0.5} \underset{(0.16)}{2.39}$ & $\cellcolor[rgb]{1,0.5,0.5} \underset{(55.29)}{57.65}$ & $\cellcolor[rgb]{0.5,1,0.5} \underset{(0.12)}{2.4}$ & $\cellcolor[rgb]{0.501,1,0.5} \underset{(0.11)}{2.44}$ & $\cellcolor[rgb]{0.506,1,0.5} \underset{(0.11)}{2.73}$ \\ 
  4 & $\cellcolor[rgb]{1,0.5,0.5} \underset{(0.23)}{4.55}$ & $\cellcolor[rgb]{0.5,1,0.5} \underset{(0.15)}{2.35}$ & $\cellcolor[rgb]{0.506,1,0.5} \underset{(0.12)}{2.36}$ & $\cellcolor[rgb]{0.568,1,0.5} \underset{(0.12)}{2.5}$ & $\cellcolor[rgb]{0.535,1,0.5} \underset{(0.09)}{2.42}$ & $\cellcolor[rgb]{0.675,1,0.5} \underset{(0.12)}{2.73}$ \\ 
  5 & $\cellcolor[rgb]{1,0.5,0.5} \underset{(0.25)}{4.63}$ & $\cellcolor[rgb]{0.5,1,0.5} \underset{(0.14)}{2.32}$ & $\cellcolor[rgb]{0.585,1,0.5} \underset{(0.12)}{2.52}$ & $\cellcolor[rgb]{0.56,1,0.5} \underset{(0.12)}{2.46}$ & $\cellcolor[rgb]{0.566,1,0.5} \underset{(0.1)}{2.47}$ & $\cellcolor[rgb]{0.818,1,0.5} \underset{(0.12)}{3.05}$ \\ 
  6 & $\cellcolor[rgb]{1,0.5,0.5} \underset{(0.24)}{4.56}$ & $\cellcolor[rgb]{0.554,1,0.5} \underset{(0.16)}{2.49}$ & $\cellcolor[rgb]{0.5,1,0.5} \underset{(0.12)}{2.37}$ & $\cellcolor[rgb]{0.518,1,0.5} \underset{(0.12)}{2.41}$ & $\cellcolor[rgb]{0.524,1,0.5} \underset{(0.12)}{2.43}$ & $\cellcolor[rgb]{0.777,1,0.5} \underset{(0.12)}{2.98}$ \\ 
  7 & $\cellcolor[rgb]{1,0.5,0.5} \underset{(0.23)}{4.48}$ & $\cellcolor[rgb]{0.683,1,0.5} \underset{(0.2)}{3.12}$ & $\cellcolor[rgb]{0.535,1,0.5} \underset{(0.19)}{2.87}$ & $\cellcolor[rgb]{0.539,1,0.5} \underset{(0.19)}{2.88}$ & $\cellcolor[rgb]{0.5,1,0.5} \underset{(0.18)}{2.82}$ & $\cellcolor[rgb]{0.828,1,0.5} \underset{(0.21)}{3.36}$ \\ 
  8 & $\cellcolor[rgb]{1,0.5,0.5} \underset{(0.25)}{4.44}$ & $\cellcolor[rgb]{0.816,1,0.5} \underset{(0.23)}{3.57}$ & $\cellcolor[rgb]{0.5,1,0.5} \underset{(0.2)}{3.16}$ & $\cellcolor[rgb]{0.757,1,0.5} \underset{(0.38)}{3.49}$ & $\cellcolor[rgb]{0.527,1,0.5} \underset{(0.21)}{3.2}$ & $\cellcolor[rgb]{0.607,1,0.5} \underset{(0.2)}{3.3}$ \\ 
  9 & $\cellcolor[rgb]{1,0.5,0.5} \underset{(0.26)}{4.58}$ & $\cellcolor[rgb]{0.954,1,0.5} \underset{(0.25)}{3.91}$ & $\cellcolor[rgb]{0.5,1,0.5} \underset{(0.22)}{3.35}$ & $\cellcolor[rgb]{0.604,1,0.5} \underset{(0.22)}{3.48}$ & $\cellcolor[rgb]{0.821,1,0.5} \underset{(0.25)}{3.74}$ & $\cellcolor[rgb]{0.551,1,0.5} \underset{(0.2)}{3.41}$ \\ 
  10 & $\cellcolor[rgb]{0.596,1,0.5} \underset{(0.26)}{4.64}$ & $\cellcolor[rgb]{1,0.5,0.5} \underset{(0.33)}{5.63}$ & $\cellcolor[rgb]{0.923,1,0.5} \underset{(0.28)}{4.99}$ & $\cellcolor[rgb]{0.996,1,0.5} \underset{(0.29)}{5.07}$ & $\cellcolor[rgb]{1,0.549,0.5} \underset{(0.33)}{5.57}$ & $\cellcolor[rgb]{0.5,1,0.5} \underset{(0.26)}{4.53}$ \\ 
  11 & $\cellcolor[rgb]{0.5,1,0.5} \underset{(0.25)}{4.6}$ & $\cellcolor[rgb]{1,0.712,0.5} \underset{(0.35)}{6.24}$ & $\cellcolor[rgb]{1,0.5,0.5} \underset{(0.35)}{6.68}$ & $\cellcolor[rgb]{1,0.535,0.5} \underset{(0.34)}{6.6}$ & $\cellcolor[rgb]{1,0.518,0.5} \underset{(0.36)}{6.64}$ & $\cellcolor[rgb]{0.909,1,0.5} \underset{(0.31)}{5.45}$ \\ 
  12 & $\cellcolor[rgb]{0.5,1,0.5} \underset{(0.26)}{4.71}$ & $\cellcolor[rgb]{1,0.5,0.5} \underset{(0.34)}{6.32}$ & $\cellcolor[rgb]{0.815,1,0.5} \underset{(0.28)}{5.22}$ & $\cellcolor[rgb]{0.766,1,0.5} \underset{(0.28)}{5.14}$ & $\cellcolor[rgb]{0.782,1,0.5} \underset{(0.32)}{5.17}$ & $\cellcolor[rgb]{0.64,1,0.5} \underset{(0.29)}{4.94}$ \\ 
  13 & $\cellcolor[rgb]{0.96,1,0.5} \underset{(0.29)}{4.89}$ & $\cellcolor[rgb]{1,0.5,0.5} \underset{(0.33)}{5.53}$ & $\cellcolor[rgb]{0.644,1,0.5} \underset{(0.24)}{4.52}$ & $\cellcolor[rgb]{0.925,1,0.5} \underset{(0.28)}{4.85}$ & $\cellcolor[rgb]{0.5,1,0.5} \underset{(0.32)}{4.35}$ & $\cellcolor[rgb]{1,0.944,0.5} \underset{(0.29)}{5.01}$ \\ 
  14 & $\cellcolor[rgb]{1,0.919,0.5} \underset{(0.3)}{4.87}$ & $\cellcolor[rgb]{1,0.56,0.5} \underset{(0.28)}{5.21}$ & $\cellcolor[rgb]{1,0.951,0.5} \underset{(0.26)}{4.84}$ & $\cellcolor[rgb]{1,0.669,0.5} \underset{(0.26)}{5.11}$ & $\cellcolor[rgb]{0.5,1,0.5} \underset{(0.3)}{4.33}$ & $\cellcolor[rgb]{1,0.5,0.5} \underset{(0.35)}{5.27}$ \\ 
  15 & $\cellcolor[rgb]{1,0.946,0.5} \underset{(0.33)}{5.04}$ & $\cellcolor[rgb]{0.884,1,0.5} \underset{(0.27)}{4.84}$ & $\cellcolor[rgb]{1,0.5,0.5} \underset{(0.31)}{5.56}$ & $\cellcolor[rgb]{1,0.733,0.5} \underset{(0.3)}{5.29}$ & $\cellcolor[rgb]{0.5,1,0.5} \underset{(0.27)}{4.39}$ & $\cellcolor[rgb]{1,0.568,0.5} \underset{(0.31)}{5.48}$ \\ 
  16 & $\cellcolor[rgb]{0.824,1,0.5} \underset{(0.3)}{5}$ & $\cellcolor[rgb]{0.705,1,0.5} \underset{(0.3)}{4.76}$ & $\cellcolor[rgb]{1,0.949,0.5} \underset{(0.27)}{5.47}$ & $\cellcolor[rgb]{0.89,1,0.5} \underset{(0.26)}{5.14}$ & $\cellcolor[rgb]{0.5,1,0.5} \underset{(0.27)}{4.34}$ & $\cellcolor[rgb]{1,0.5,0.5} \underset{(0.43)}{6.4}$ \\ 
  17 & $\cellcolor[rgb]{1,0.876,0.5} \underset{(0.3)}{5}$ & $\cellcolor[rgb]{0.762,1,0.5} \underset{(0.26)}{4.22}$ & $\cellcolor[rgb]{1,0.996,0.5} \underset{(0.26)}{4.74}$ & $\cellcolor[rgb]{0.939,1,0.5} \underset{(0.26)}{4.6}$ & $\cellcolor[rgb]{0.5,1,0.5} \underset{(0.24)}{3.65}$ & $\cellcolor[rgb]{1,0.5,0.5} \underset{(0.5)}{5.82}$ \\ 
  18 & $\cellcolor[rgb]{1,0.5,0.5} \underset{(0.34)}{5.19}$ & $\cellcolor[rgb]{0.989,1,0.5} \underset{(0.28)}{4.28}$ & $\cellcolor[rgb]{0.5,1,0.5} \underset{(0.19)}{3.4}$ & $\cellcolor[rgb]{0.575,1,0.5} \underset{(0.21)}{3.53}$ & $\cellcolor[rgb]{0.513,1,0.5} \underset{(0.25)}{3.42}$ & $\cellcolor[rgb]{0.738,1,0.5} \underset{(0.21)}{3.83}$ \\ 
  19 & $\cellcolor[rgb]{1,0.5,0.5} \underset{(0.33)}{4.98}$ & $\cellcolor[rgb]{0.962,1,0.5} \underset{(0.3)}{4.2}$ & $\cellcolor[rgb]{0.586,1,0.5} \underset{(0.2)}{3.65}$ & $\cellcolor[rgb]{0.605,1,0.5} \underset{(0.22)}{3.68}$ & $\cellcolor[rgb]{0.602,1,0.5} \underset{(0.25)}{3.67}$ & $\cellcolor[rgb]{0.5,1,0.5} \underset{(0.15)}{3.52}$ \\ 
  20 & $\cellcolor[rgb]{1,0.803,0.5} \underset{(0.31)}{4.88}$ & $\cellcolor[rgb]{1,0.5,0.5} \underset{(0.31)}{5.44}$ & $\cellcolor[rgb]{0.612,1,0.5} \underset{(0.21)}{3.79}$ & $\cellcolor[rgb]{0.616,1,0.5} \underset{(0.21)}{3.8}$ & $\cellcolor[rgb]{0.612,1,0.5} \underset{(0.22)}{3.79}$ & $\cellcolor[rgb]{0.5,1,0.5} \underset{(0.17)}{3.59}$ \\ 
  21 & $\cellcolor[rgb]{1,0.5,0.5} \underset{(0.29)}{4.73}$ & $\cellcolor[rgb]{0.57,1,0.5} \underset{(0.17)}{3.07}$ & $\cellcolor[rgb]{0.541,1,0.5} \underset{(0.13)}{3.02}$ & $\cellcolor[rgb]{0.5,1,0.5} \underset{(0.13)}{2.95}$ & $\cellcolor[rgb]{0.599,1,0.5} \underset{(0.15)}{3.13}$ & $\cellcolor[rgb]{0.582,1,0.5} \underset{(0.13)}{3.1}$ \\ 
  22 & $\cellcolor[rgb]{1,0.5,0.5} \underset{(0.27)}{4.31}$ & $\cellcolor[rgb]{0.938,1,0.5} \underset{(0.18)}{3.32}$ & $\cellcolor[rgb]{0.5,1,0.5} \underset{(0.12)}{2.55}$ & $\cellcolor[rgb]{0.534,1,0.5} \underset{(0.13)}{2.61}$ & $\cellcolor[rgb]{0.59,1,0.5} \underset{(0.15)}{2.71}$ & $\cellcolor[rgb]{0.925,1,0.5} \underset{(0.15)}{3.3}$ \\ 
  23 & $\cellcolor[rgb]{1,0.5,0.5} \underset{(0.3)}{4.35}$ & $\cellcolor[rgb]{0.948,1,0.5} \underset{(0.19)}{3.32}$ & $\cellcolor[rgb]{0.537,1,0.5} \underset{(0.15)}{2.55}$ & $\cellcolor[rgb]{0.5,1,0.5} \underset{(0.15)}{2.48}$ & $\cellcolor[rgb]{0.589,1,0.5} \underset{(0.19)}{2.65}$ & $\cellcolor[rgb]{0.779,1,0.5} \underset{(0.15)}{3}$ \\ 
  24 & $\cellcolor[rgb]{1,0.5,0.5} \underset{(0.27)}{4.4}$ & $\cellcolor[rgb]{0.531,1,0.5} \underset{(0.16)}{3.04}$ & $\cellcolor[rgb]{0.579,1,0.5} \underset{(0.14)}{3.11}$ & $\cellcolor[rgb]{0.712,1,0.5} \underset{(0.16)}{3.29}$ & $\cellcolor[rgb]{0.5,1,0.5} \underset{(0.15)}{3}$ & $\cellcolor[rgb]{0.635,1,0.5} \underset{(0.13)}{3.19}$ \\ 
   \hline
\end{tabular}
\begin{tablenotes}
\footnotesize
\item Note: NO, JSU, and SHASH denote the Normal, Johnson's SU, and Sinh–Arcsinh distributions, respectively. The number in parentheses next to each distribution name indicates how many distribution moments are modelled as non-constant, i.e., as functions of the explanatory variables. For example, JSU(2) means that the first two moments are non-constant. Parentheses beneath the error measures contain the corresponding standard deviations. A heat map is used to indicate lower (green) and higher (red) forecast errors.
\end{tablenotes}
\end{threeparttable}
\end{table}

\newpage

\begin{table}[H]
\centering
\caption{PB $\tau=0.01$ by distribution and hour}\label{tab:PB1_hour}
\hspace{-3cm}
\begin{threeparttable}
\begin{tabular}{rcccccc}
  \hline
Hour & Naive & NO & JSU(2) & JSU(3) & JSU(4) & SHASH(2) \\ 
  \hline
1 & $\cellcolor[rgb]{1,0.5,0.5} \underset{(0.19)}{1.65}$ & $\cellcolor[rgb]{0.744,1,0.5} \underset{(0.12)}{0.9}$ & $\cellcolor[rgb]{0.5,1,0.5} \underset{(0.06)}{0.66}$ & $\cellcolor[rgb]{0.518,1,0.5} \underset{(0.06)}{0.68}$ & $\cellcolor[rgb]{0.532,1,0.5} \underset{(0.02)}{0.69}$ & $\cellcolor[rgb]{0.607,1,0.5} \underset{(0.07)}{0.77}$ \\ 
  2 & $\cellcolor[rgb]{1,0.5,0.5} \underset{(0.15)}{1.47}$ & $\cellcolor[rgb]{0.844,1,0.5} \underset{(0.12)}{0.99}$ & $\cellcolor[rgb]{0.5,1,0.5} \underset{(0.06)}{0.74}$ & $\cellcolor[rgb]{0.652,1,0.5} \underset{(0.08)}{0.85}$ & $\cellcolor[rgb]{0.547,1,0.5} \underset{(0.03)}{0.77}$ & $\cellcolor[rgb]{0.654,1,0.5} \underset{(0.06)}{0.85}$ \\ 
  3 & $\cellcolor[rgb]{0.512,1,0.5} \underset{(0.15)}{1.42}$ & $\cellcolor[rgb]{0.504,1,0.5} \underset{(0.12)}{0.98}$ & $\cellcolor[rgb]{1,0.5,0.5} \underset{(56.4)}{57.16}$ & $\cellcolor[rgb]{0.5,1,0.5} \underset{(0.05)}{0.76}$ & $\cellcolor[rgb]{0.503,1,0.5} \underset{(0.03)}{0.91}$ & $\cellcolor[rgb]{0.502,1,0.5} \underset{(0.05)}{0.87}$ \\ 
  4 & $\cellcolor[rgb]{1,0.5,0.5} \underset{(0.15)}{1.47}$ & $\cellcolor[rgb]{0.737,1,0.5} \underset{(0.11)}{0.9}$ & $\cellcolor[rgb]{0.5,1,0.5} \underset{(0.05)}{0.73}$ & $\cellcolor[rgb]{0.569,1,0.5} \underset{(0.04)}{0.78}$ & $\cellcolor[rgb]{0.756,1,0.5} \underset{(0.03)}{0.92}$ & $\cellcolor[rgb]{0.657,1,0.5} \underset{(0.05)}{0.84}$ \\ 
  5 & $\cellcolor[rgb]{1,0.5,0.5} \underset{(0.17)}{1.55}$ & $\cellcolor[rgb]{0.632,1,0.5} \underset{(0.1)}{0.88}$ & $\cellcolor[rgb]{0.5,1,0.5} \underset{(0.04)}{0.77}$ & $\cellcolor[rgb]{0.501,1,0.5} \underset{(0.04)}{0.78}$ & $\cellcolor[rgb]{0.657,1,0.5} \underset{(0.02)}{0.9}$ & $\cellcolor[rgb]{0.779,1,0.5} \underset{(0.05)}{0.99}$ \\ 
  6 & $\cellcolor[rgb]{1,0.5,0.5} \underset{(0.17)}{1.49}$ & $\cellcolor[rgb]{0.9,1,0.5} \underset{(0.12)}{1.02}$ & $\cellcolor[rgb]{0.5,1,0.5} \underset{(0.05)}{0.71}$ & $\cellcolor[rgb]{0.543,1,0.5} \underset{(0.05)}{0.74}$ & $\cellcolor[rgb]{0.617,1,0.5} \underset{(0.04)}{0.8}$ & $\cellcolor[rgb]{0.831,1,0.5} \underset{(0.05)}{0.97}$ \\ 
  7 & $\cellcolor[rgb]{1,0.5,0.5} \underset{(0.15)}{1.46}$ & $\cellcolor[rgb]{1,0.621,0.5} \underset{(0.15)}{1.38}$ & $\cellcolor[rgb]{0.6,1,0.5} \underset{(0.09)}{0.88}$ & $\cellcolor[rgb]{0.574,1,0.5} \underset{(0.08)}{0.86}$ & $\cellcolor[rgb]{0.5,1,0.5} \underset{(0.06)}{0.82}$ & $\cellcolor[rgb]{0.814,1,0.5} \underset{(0.1)}{1.02}$ \\ 
  8 & $\cellcolor[rgb]{1,0.525,0.5} \underset{(0.19)}{1.48}$ & $\cellcolor[rgb]{1,0.5,0.5} \underset{(0.17)}{1.5}$ & $\cellcolor[rgb]{0.684,1,0.5} \underset{(0.11)}{0.9}$ & $\cellcolor[rgb]{0.8,1,0.5} \underset{(0.13)}{0.99}$ & $\cellcolor[rgb]{0.5,1,0.5} \underset{(0.08)}{0.76}$ & $\cellcolor[rgb]{0.838,1,0.5} \underset{(0.12)}{1.01}$ \\ 
  9 & $\cellcolor[rgb]{1,0.657,0.5} \underset{(0.19)}{1.58}$ & $\cellcolor[rgb]{1,0.5,0.5} \underset{(0.19)}{1.68}$ & $\cellcolor[rgb]{0.52,1,0.5} \underset{(0.13)}{1.06}$ & $\cellcolor[rgb]{0.532,1,0.5} \underset{(0.12)}{1.07}$ & $\cellcolor[rgb]{0.646,1,0.5} \underset{(0.12)}{1.14}$ & $\cellcolor[rgb]{0.5,1,0.5} \underset{(0.11)}{1.05}$ \\ 
  10 & $\cellcolor[rgb]{0.571,1,0.5} \underset{(0.19)}{1.57}$ & $\cellcolor[rgb]{1,0.5,0.5} \underset{(0.26)}{3.07}$ & $\cellcolor[rgb]{0.688,1,0.5} \underset{(0.17)}{1.76}$ & $\cellcolor[rgb]{0.708,1,0.5} \underset{(0.17)}{1.79}$ & $\cellcolor[rgb]{0.752,1,0.5} \underset{(0.19)}{1.86}$ & $\cellcolor[rgb]{0.5,1,0.5} \underset{(0.15)}{1.45}$ \\ 
  11 & $\cellcolor[rgb]{0.5,1,0.5} \underset{(0.19)}{1.52}$ & $\cellcolor[rgb]{1,0.5,0.5} \underset{(0.29)}{3.4}$ & $\cellcolor[rgb]{1,0.883,0.5} \underset{(0.23)}{2.68}$ & $\cellcolor[rgb]{1,0.954,0.5} \underset{(0.22)}{2.55}$ & $\cellcolor[rgb]{0.966,1,0.5} \underset{(0.21)}{2.4}$ & $\cellcolor[rgb]{0.718,1,0.5} \underset{(0.19)}{1.93}$ \\ 
  12 & $\cellcolor[rgb]{0.55,1,0.5} \underset{(0.19)}{1.6}$ & $\cellcolor[rgb]{1,0.5,0.5} \underset{(0.26)}{3.34}$ & $\cellcolor[rgb]{0.523,1,0.5} \underset{(0.14)}{1.55}$ & $\cellcolor[rgb]{0.589,1,0.5} \underset{(0.15)}{1.67}$ & $\cellcolor[rgb]{0.5,1,0.5} \underset{(0.17)}{1.51}$ & $\cellcolor[rgb]{0.579,1,0.5} \underset{(0.17)}{1.65}$ \\ 
  13 & $\cellcolor[rgb]{0.815,1,0.5} \underset{(0.22)}{1.76}$ & $\cellcolor[rgb]{1,0.5,0.5} \underset{(0.26)}{2.83}$ & $\cellcolor[rgb]{0.528,1,0.5} \underset{(0.1)}{1.32}$ & $\cellcolor[rgb]{0.545,1,0.5} \underset{(0.11)}{1.34}$ & $\cellcolor[rgb]{0.5,1,0.5} \underset{(0.17)}{1.27}$ & $\cellcolor[rgb]{0.664,1,0.5} \underset{(0.11)}{1.53}$ \\ 
  14 & $\cellcolor[rgb]{0.772,1,0.5} \underset{(0.24)}{1.75}$ & $\cellcolor[rgb]{1,0.5,0.5} \underset{(0.22)}{2.73}$ & $\cellcolor[rgb]{0.702,1,0.5} \underset{(0.13)}{1.66}$ & $\cellcolor[rgb]{0.773,1,0.5} \underset{(0.14)}{1.75}$ & $\cellcolor[rgb]{0.5,1,0.5} \underset{(0.17)}{1.39}$ & $\cellcolor[rgb]{0.819,1,0.5} \underset{(0.15)}{1.82}$ \\ 
  15 & $\cellcolor[rgb]{1,0.935,0.5} \underset{(0.26)}{1.9}$ & $\cellcolor[rgb]{1,0.5,0.5} \underset{(0.2)}{2.39}$ & $\cellcolor[rgb]{0.932,1,0.5} \underset{(0.13)}{1.76}$ & $\cellcolor[rgb]{0.997,1,0.5} \underset{(0.14)}{1.83}$ & $\cellcolor[rgb]{0.5,1,0.5} \underset{(0.12)}{1.27}$ & $\cellcolor[rgb]{0.875,1,0.5} \underset{(0.12)}{1.69}$ \\ 
  16 & $\cellcolor[rgb]{0.854,1,0.5} \underset{(0.23)}{1.81}$ & $\cellcolor[rgb]{1,0.5,0.5} \underset{(0.24)}{2.55}$ & $\cellcolor[rgb]{0.835,1,0.5} \underset{(0.12)}{1.78}$ & $\cellcolor[rgb]{0.77,1,0.5} \underset{(0.12)}{1.71}$ & $\cellcolor[rgb]{0.5,1,0.5} \underset{(0.14)}{1.4}$ & $\cellcolor[rgb]{0.965,1,0.5} \underset{(0.16)}{1.93}$ \\ 
  17 & $\cellcolor[rgb]{1,0.665,0.5} \underset{(0.23)}{1.82}$ & $\cellcolor[rgb]{1,0.5,0.5} \underset{(0.2)}{1.93}$ & $\cellcolor[rgb]{0.818,1,0.5} \underset{(0.11)}{1.45}$ & $\cellcolor[rgb]{0.691,1,0.5} \underset{(0.1)}{1.36}$ & $\cellcolor[rgb]{0.5,1,0.5} \underset{(0.12)}{1.23}$ & $\cellcolor[rgb]{1,0.642,0.5} \underset{(0.18)}{1.83}$ \\ 
  18 & $\cellcolor[rgb]{1,0.5,0.5} \underset{(0.27)}{2.02}$ & $\cellcolor[rgb]{1,0.742,0.5} \underset{(0.21)}{1.77}$ & $\cellcolor[rgb]{0.5,1,0.5} \underset{(0.07)}{1.01}$ & $\cellcolor[rgb]{0.569,1,0.5} \underset{(0.1)}{1.08}$ & $\cellcolor[rgb]{0.506,1,0.5} \underset{(0.1)}{1.01}$ & $\cellcolor[rgb]{0.713,1,0.5} \underset{(0.1)}{1.22}$ \\ 
  19 & $\cellcolor[rgb]{1,0.5,0.5} \underset{(0.27)}{1.97}$ & $\cellcolor[rgb]{1,0.607,0.5} \underset{(0.24)}{1.87}$ & $\cellcolor[rgb]{0.5,1,0.5} \underset{(0.09)}{0.98}$ & $\cellcolor[rgb]{0.587,1,0.5} \underset{(0.12)}{1.07}$ & $\cellcolor[rgb]{0.619,1,0.5} \underset{(0.13)}{1.1}$ & $\cellcolor[rgb]{0.502,1,0.5} \underset{(0.05)}{0.99}$ \\ 
  20 & $\cellcolor[rgb]{0.844,1,0.5} \underset{(0.25)}{1.85}$ & $\cellcolor[rgb]{1,0.5,0.5} \underset{(0.28)}{3.14}$ & $\cellcolor[rgb]{0.576,1,0.5} \underset{(0.11)}{1.32}$ & $\cellcolor[rgb]{0.602,1,0.5} \underset{(0.12)}{1.37}$ & $\cellcolor[rgb]{0.557,1,0.5} \underset{(0.11)}{1.28}$ & $\cellcolor[rgb]{0.5,1,0.5} \underset{(0.09)}{1.17}$ \\ 
  21 & $\cellcolor[rgb]{1,0.5,0.5} \underset{(0.22)}{1.74}$ & $\cellcolor[rgb]{0.888,1,0.5} \underset{(0.14)}{1.27}$ & $\cellcolor[rgb]{0.56,1,0.5} \underset{(0.08)}{1.02}$ & $\cellcolor[rgb]{0.504,1,0.5} \underset{(0.07)}{0.98}$ & $\cellcolor[rgb]{0.5,1,0.5} \underset{(0.05)}{0.98}$ & $\cellcolor[rgb]{0.626,1,0.5} \underset{(0.09)}{1.07}$ \\ 
  22 & $\cellcolor[rgb]{1,0.5,0.5} \underset{(0.21)}{1.65}$ & $\cellcolor[rgb]{1,0.579,0.5} \underset{(0.15)}{1.58}$ & $\cellcolor[rgb]{0.547,1,0.5} \underset{(0.08)}{0.9}$ & $\cellcolor[rgb]{0.626,1,0.5} \underset{(0.09)}{0.96}$ & $\cellcolor[rgb]{0.5,1,0.5} \underset{(0.06)}{0.86}$ & $\cellcolor[rgb]{0.886,1,0.5} \underset{(0.08)}{1.17}$ \\ 
  23 & $\cellcolor[rgb]{1,0.5,0.5} \underset{(0.24)}{1.71}$ & $\cellcolor[rgb]{1,0.59,0.5} \underset{(0.16)}{1.63}$ & $\cellcolor[rgb]{0.545,1,0.5} \underset{(0.08)}{0.88}$ & $\cellcolor[rgb]{0.5,1,0.5} \underset{(0.08)}{0.84}$ & $\cellcolor[rgb]{0.544,1,0.5} \underset{(0.08)}{0.88}$ & $\cellcolor[rgb]{0.753,1,0.5} \underset{(0.08)}{1.06}$ \\ 
  24 & $\cellcolor[rgb]{1,0.5,0.5} \underset{(0.2)}{1.65}$ & $\cellcolor[rgb]{0.786,1,0.5} \underset{(0.11)}{1.12}$ & $\cellcolor[rgb]{0.555,1,0.5} \underset{(0.07)}{0.94}$ & $\cellcolor[rgb]{0.62,1,0.5} \underset{(0.07)}{0.99}$ & $\cellcolor[rgb]{0.5,1,0.5} \underset{(0.04)}{0.9}$ & $\cellcolor[rgb]{0.547,1,0.5} \underset{(0.06)}{0.94}$ \\ 
   \hline
\end{tabular}
\begin{tablenotes}
\footnotesize
\item Note: NO, JSU, and SHASH denote the Normal, Johnson's SU, and Sinh–Arcsinh distributions, respectively. The number in parentheses next to each distribution name indicates how many distribution moments are modelled as non-constant, i.e., as functions of the explanatory variables. For example, JSU(2) means that the first two moments are non-constant. Parentheses beneath the error measures contain the corresponding standard deviations. A heat map is used to indicate lower (green) and higher (red) forecast errors.
\end{tablenotes}
\end{threeparttable}
\end{table}

\end{appendices}

\end{document}